\documentclass[a4paper]{IEEEtran}

\usepackage{physics}
\usepackage[T1]{fontenc}
\usepackage[utf8]{inputenc}
\usepackage[english]{babel}
\usepackage[active]{srcltx}
\usepackage{color}
\usepackage{amsmath}
\usepackage{amssymb}
\usepackage[final]{graphicx}
\usepackage{amsbsy}
\usepackage{url}
\usepackage{enumerate}
\usepackage{mathrsfs}
\usepackage{latexsym}
\usepackage{cite}
\usepackage{amsfonts}
\usepackage{booktabs}
\usepackage{tabularx}
\usepackage{graphicx}
\usepackage{textcomp}
\usepackage{xcolor}
\usepackage{verbatim}

\usepackage{stfloats}
\usepackage{dsfont}
\usepackage{gensymb}    
\usepackage{booktabs}   
\usepackage{comment}    
\usepackage{euscript}

\usepackage{amsthm}
\theoremstyle{definition}
\newtheorem{remark}{Remark}
\AtEndEnvironment{remark}{\hfill$\blacksquare$}

\def\be#1\ee{\begin{equation}#1\end{equation}}
\def\barr#1\earr{\begin{align}#1\end{align}}

\newcommand{\eqdef}{\triangleq}
\newcommand{\bm}[1]{{\mathbf{#1}}}

\newcommand{\Tu}{T_{\text{u}}}
\newcommand{\Tcp}{T_{\text{cp}}}
\newcommand{\Tcoh}{T_{\text{coh}}}
\renewcommand{\Tr}{T_{\text{r}}}
\newcommand{\Ts}{T_{\text{s}}}
\newcommand{\Tc}{T_{\text{c}}}
\newcommand{\Bs}{B_{\text{s}}}
\newcommand{\Nc}{N_{\text{c}}}
\newcommand{\Np}{N_{\text{p}}}
\newcommand{\Nz}{N_{\text{z}}}
\newcommand{\Qc}{Q_{\text{c}}}

\newcommand{\Ls}{L_{\text{s}}}
\newcommand{\Lcoh}{L_{\text{coh}}}
\newcommand{\ku}{\eta_{\text{u}}}
\newcommand{\pil}{\text{pil}}

\newcommand{\infint}{\int_{-\infty}^{+\infty}}
\newcommand{\herm}{\text{H}}
\newcommand{\trasp}{\text{T}}
\newcommand{\Cset}{\mathbb{C}}
\newcommand{\Rset}{\mathbb{R}}
\newcommand{\Nset}{\mathbb{N}}
\newcommand{\Zset}{\mathbb{Z}}

\newcommand{\diag}{\text{diag}}
\newcommand{\sinc}{\text{sinc}}
\newcommand{\Es}{{\mathbb{E}}}

\newcommand{\BP}{\text{bp}}

\newcommand{\thetaout}{\theta_\text{r}}
\newcommand{\thetain}{\theta_{\text{s}}}

\newcommand{\betahat}{\widehat{\beta}}
\newcommand{\Tgamma}{T_{\gamma}}

\newcounter{assumption}
\renewcommand{\theassumption}{a\arabic{assumption}}
\newcommand{\assumption}[2][]{%
    \refstepcounter{assumption}%
    \ifx\relax#1\relax\else\label{#1}\fi
    \textbf{(\theassumption)}~#2%
}

\newcounter{temp_counter}

\begin{document}


\title{A Dual-Mode Protocol for Space-Time-Coding Metasurfaces in OFDM ISAC Systems}

\author{
Donatella~Darsena,~\IEEEmembership{Senior Member,~IEEE},
Ivan~Iudice,~\IEEEmembership{Senior Member,~IEEE},
Giacinto~Gelli,~\IEEEmembership{Senior Member,~IEEE},
and~Vincenzo~Galdi,~\IEEEmembership{Fellow,~IEEE}
}

\maketitle

\begin{abstract}
Existing integrated sensing and communication (ISAC) frameworks assisted by
space-time-coding metasurfaces (STCMs) rely on idealized surface models,
which neglect the frequency-selective and angle-dependent response of the
meta-atoms, and are typically limited to narrowband single-carrier
communications.
This paper addresses these limitations by developing a physics-based
signal-processing framework for STCM-assisted ISAC systems employing 
wideband orthogonal frequency-division
multiplexing (OFDM) waveforms, which accounts
also for beam squint effects across the STCM aperture. 
On this basis, we propose a dual-mode protocol, in which the STCM operates
as a time-varying surface during a short acquisition phase and as a
conventional reconfigurable intelligent surface (RIS) during data
transmission.
In the acquisition phase, the angle of arrival (AoA) 
and the {composite} channel
are jointly estimated 
from a single pilot OFDM
symbol transmitted over a subset of
active subcarriers, together with deliberately introduced virtual carriers; 
these estimates are then used to configure the
RIS for data transmission.
Numerical results validate the 
considered framework
and quantify the gains 
achievable by the proposed 
dual-mode protocol. 
\end{abstract}



\section{Introduction}
\label{sec:intro}

\IEEEPARstart{I}{ntegrated} sensing and communication (ISAC) has emerged as
one of the key enabling technologies for next-generation wireless
networks~\cite{Zhang2026}, driven by applications requiring both high-rate
communications and accurate environmental sensing.
%
By sharing hardware, spectrum, and waveforms, 
ISAC improves spectral efficiency and hardware utilization 
while reducing deployment costs and
energy consumption~\cite{Hu2024}.
\textit{Reconfigurable intelligent surfaces} (RISs) 
constitute another enabling
technology for beyond-5G (B5G) and 6G wireless systems. 
They comprise large arrays
of nearly passive subwavelength elements (``meta-atoms''), whose
electromagnetic (EM) response, in terms of phase and possibly amplitude, 
can be dynamically controlled to reconfigure
the propagation environment without requiring 
active radio-frequency (RF) chains.
When employed in ISAC, RISs can create 
virtual line-of-sight (LoS) links and provide
additional spatial degrees of freedom (DoFs), thereby mitigating path loss
and blockage and enhancing both sensing and communication
performance~\cite{Liu2023}.

\textit{Space-time-coding metasurfaces} (STCMs)~\cite{Zhang2018}, which extend the coding metasurfaces of~\cite{Cui2014} and are also referred to as \textit{space-time-coded RISs} (STC-RISs), add controlled temporal modulation to spatial reconfiguration.
Periodic modulation or \textit{coding} of the meta-atoms, typically implemented 
with low-complexity hardware
based on PIN or varactor diodes, generates harmonics
around the carrier frequency. Joint design of 
the spatial and temporal coding patterns enables simultaneous control of the fundamental
component and of the harmonics, thereby providing additional DoFs for beam
steering, waveform shaping, and 
multiplexing.

STCMs are particularly attractive for ISAC,
since the same temporal DoFs can be exploited for target detection,
localization, and parameter estimation, while the above communication
functionalities of RISs are retained~\cite{Chen2024,Chen2025}.
Their integration with orthogonal frequency-division multiplexing (OFDM) is
especially appealing. By choosing the STCM modulation period such that its
fundamental frequency equals the OFDM subcarrier spacing, 
harmonic components generated by the temporal modulation 
turn out to be aligned with the OFDM
frequency grid. Subject to suitable cyclic prefix (CP) conditions, these
components can be extracted through conventional OFDM demodulation and
exploited as additional {virtual observations} 
of the incident signal.



\subsection{Related work}
\label{ssec:related}

Dos Santos \textit{et al.}~\cite{Santos2024} exploited the space-frequency
scattering induced by periodic switching of the metasurface elements for
target detection, localization, and classification, turning the frequency
conversion of STCMs from a communication impairment into a sensing
resource.
Chen \textit{et al.}~\cite{Chen2025} validated a
fully integrated STCM-assisted ISAC architecture, in which the fundamental
frequency supports the communication link while harmonic beams are used for
sensing, and demonstrated 
sub-degree angle-of-arrival (AoA) estimation on a
2-bit prototype working at 10.3~GHz.
STCM prototypes for concurrent beam steering and data
modulation~\cite{Gholami2026} and unified frameworks for metasurface-enabled
ISAC transceivers~\cite{Li2026} have also been developed.

Recent studies have also explored the use of STCM-generated harmonics for
angular estimation. Chen \textit{et al.}~\cite{Chen2026} estimated the AoA
from the complex correlations among the harmonic components generated by an
STCM, under a narrowband, frequency-flat, and angle-independent meta-atom
model. 
Hu \textit{et al.}~\cite{Hu2025} focused instead on the 
{design} of the coding sequence that maximizes 
AoA estimation accuracy for a \textit{transmissive} RIS observed 
by a single receiving sensor. 
A different architecture was considered in \cite{Mizmizi2026}, 
where an STC-RIS conveys uplink information 
by imposing continuous phase modulation on the
reflected radar waveform.
These approaches depart from classical AoA estimation, in which sensor
arrays process multiple snapshots of the received
signals~\cite{Krim1996}, e.g., through subspace-based techniques such as
MUSIC~\cite{Schmidt1986} and ESPRIT~\cite{Roy1989}.
RIS-based schemes~\cite{Chen2021,Yu2025} occupy an intermediate position:
an RIS whose reflection coefficients are reconfigured over multiple time slots
acts as a \textit{virtual array}, whose observations embed the angular
parameters of the composite channel~\cite{He2020}.
These schemes generally adopt a narrowband model, in which the spatial
response is assumed to be frequency independent.

The same time-sequential reconfiguration was extended to
wideband RIS-assisted OFDM channel estimation by Ma \textit{et al.} 
in~\cite{Ma2021}.
Their model accounts for \textit{beam-squint} effects, 
i.e., the frequency dependence of the spatial phase 
progression across an electrically large aperture.
However, the RIS elements are modeled in~\cite{Ma2021}
as ideal phase shifters, whose responses depend 
neither on frequency nor on the incidence angle.
Moreover, multiple OFDM training symbols 
are required to reconfigure the RIS.
In contrast, an STCM generates multiple 
harmonic observations across
\textit{frequency} 
within a single training OFDM symbol, 
thereby requiring
neither reconfigurations over multiple OFDM symbols nor multiple receive
antennas.

\subsection{Motivation, main contributions and paper organization}
\label{ssec:contributions}

From the analysis of the literature,
two aspects remain largely unexplored in STCM-based ISAC.
First, existing models are formulated for narrowband single-carrier
signals, leaving the interaction between temporal modulation and
frequency-selective multipath propagation, and the associated broadband
channel acquisition, unaddressed.
Second, the frequency selectivity and the angular dependence of the
meta-atom response are neglected, although both may significantly affect
the response in a broadband OFDM scenario when the 
direction of incidence of the signal is unknown.


Motivated by these considerations, 
we develop in this paper a signal-processing framework for \textit{dual-mode} STCM-assisted ISAC.
A single-antenna OFDM TX communicates with a single-antenna RX through an
STCM, with the direct link blocked or severely attenuated.
Within each channel coherence interval, the STCM plays a dual role: it
operates as a time-varying RIS during a short
\textit{sensing/acquisition phase}, in which the AoA of the incident signal
and the composite propagation channel are jointly estimated, and as a
conventional time-invariant RIS during the subsequent
\textit{communication phase}.

The main contributions of the paper 
can be summarized as follows.
\begin{enumerate}

\item \textit{Physics-based meta-atom model}: We develop a \textit{linear
periodically time-varying} (LPTV) model of the meta-atom, based on a
pole--residue representation of its reflection coefficient, which jointly
accounts for temporal modulation, frequency selectivity, spatial
dispersion, and polarization.
It generalizes our previous work~\cite{Iudice2026} to a generic meta-atom
structure and recovers time-invariant, frequency-flat, and
angle-independent models as special cases.

\item \textit{STCM--OFDM processing}: We derive the input--output relation of the STCM-assisted OFDM link,
including beam-squint effects, and show that, when the STCM modulation frequency matches
the subcarrier spacing, the generated harmonics 
fall on the OFDM grid and
can be extracted at a single-antenna RX through conventional OFDM
demodulation.

\item \textit{Dual-mode ISAC protocol}: 
We propose a protocol in which the STCM operates in time-varying mode
during acquisition and in static mode during data transmission.
During acquisition, 
a single pilot OFDM symbol is transmitted over 
 a subset of subcarriers,
and the deliberately introduced 
\textit{virtual carriers}
(VCs) reduce the number of unknown channel coefficients and
enable their joint maximum-likelihood (ML) 
estimation with the AoA.
During data transmission, 
the estimated AoA allows one to configure 
the static STCM coefficients according 
to a broadband \textit{max--min criterion},
while the reconstructed channel 
enables OFDM 
equalization at the receiver.

\end{enumerate}

\medskip

\noindent
The remainder of the paper is organized as follows.
Section~\ref{sec:meta-atom-model} introduces the physics-based model of the 
meta-atoms, 
while Section~\ref{sec:system-model} describes 
the STCM-assisted ISAC system employing OFDM waveforms. 
Section~\ref{sec:dual-mode-protocol} presents the
dual-mode protocol, including joint AoA and channel estimation, and the
design of the STCM reflection coefficients. 
Numerical results and conclusions 
are given in Sections~\ref{sec:simulations} 
and~\ref{sec:conclusions},
respectively.\footnote{%
\textit{Main notations}: $\Nset\eqdef\{0,1,2,\ldots\}$,
$\mathcal{J}_N\eqdef\{1,\ldots,N\}$, and
$\mathcal{O}_N\eqdef\{0,\ldots,N-1\}$;
$\text{mod}(a,b)\eqdef a-b\lfloor a/b\rfloor\in[0,b)$ for $b>0$, and
$\arg(z)\in(-\pi,\pi]$ is the principal phase of $z\neq0$;
$(\cdot)^{\dagger}$ denotes the Moore--Penrose pseudoinverse, and
$\text{Vand}_N(\bm{z})\in\Cset^{N\times L}$ the Vandermonde matrix with
$(i,\ell)$-th entry $z_\ell^{\,i-1}$;
$\text{eig}(\bm{A})$, $\text{diag}(\bm{a})$, and $\text{range}(\bm{A})$
denote the eigenvalues, the diagonal matrix with entries $\bm{a}$, and the
column space, respectively;
$\mathcal{F}[x(t)]\eqdef\int_{-\infty}^{+\infty}x(t)\,e^{-j 2 \pi f t}\,dt$ denotes the
Fourier transform of $x(t)$;
$\text{sinc}(t)\eqdef\sin(\pi t)/(\pi t)$; $\text{u}_T(t)$ is the
rectangular pulse equal to $1$ for $t\in(0,T)$, with $\text{u}_\infty(t)$
the unit step;
$\text{rep}_T[x(t)]\eqdef\sum_{n}x(t-nT)$; and
$\mathcal{CN}(\boldsymbol{\mu},\boldsymbol{\Sigma})$ denotes the complex
proper Gaussian distribution.}


\section{Physics-based meta-atom model}
\label{sec:meta-atom-model}

The input--output model developed herein
is built upon~\cite{Iudice2026} and the 
equivalent-circuit representation
of~\cite{Costa2021}.
The meta-atom is characterized in the Laplace domain and its frequency
response is obtained by setting $s=j2\pi f$.
Under the \textit{adiabatic approximation} \cite{Minkov2017Exact}, 
the temporal modulation of the STCM is assumed to be 
slowly varying relative to the intrinsic  dynamics of the meta-atom.
Consequently, the meta-atom evolves 
through a sequence of \textit{quasi-static} states, each
characterized by its
instantaneous circuit parameters.
Let $\nu\in\{\text{TE},\text{TM}\}$ denote the polarization of the
incident EM wave and $\thetain$ its AoA, measured from the surface normal
in a two-dimensional geometry, so that no dependence on the azimuth angle
arises.
At complex frequency $s$, the $q$-th meta-atom ($q\in\mathcal{J}_Q$)
is described by its time-varying admittance
$Y_q^{(\nu)}(s,t;\thetain)$~\cite{Costa2021,Iudice2026},%
\footnote{The unit cell is assumed to exhibit no polarization coupling, so
that the two polarizations can be treated independently.}
and its bandpass reflection coefficient is
\be
\breve{\Gamma}^{(\nu)}_{\BP,q}(s,t;\thetain)
=
\frac{1-Z^{(\nu)}_0(\thetain)\,Y^{(\nu)}_q(s,t;\thetain)}
{1+Z^{(\nu)}_0(\thetain)\,Y^{(\nu)}_q(s,t;\thetain)},
\label{eq:meta-atom-reflection}
\ee
where $Z_0^{(\text{TE})}(\thetain)=\eta_0/\cos\thetain$ and
$Z_0^{(\text{TM})}(\thetain)=\eta_0\cos\thetain$ are the free-space wave
impedances~\cite{Felsen1973}, with $\eta_0\approx377~\Omega$.

In the conventional normal-incidence model
(see, e.g.,~\cite{Luukkonen:2008sa}), $\thetain=0$, so that the TE and
TM free-space wave impedances coincide, yielding identical reflection
coefficients in~\eqref{eq:meta-atom-reflection} for a
polarization-symmetric unit cell.
In general, however, the meta-atom response depends on $\thetain$, both through the free-space wave impedance and through the angular dependence of the meta-atom admittance, i.e., \textit{spatial dispersion}~\cite{Costa2021},
which cannot be disregarded here, since the AoA is the sensing parameter to
be estimated. 
For this reason, we retain $Y_q^{(\nu)}(s,t;\thetain)$ in a general form,
accounting for frequency selectivity, angular and polarization
dependence, and time reconfigurability.
Such dependencies are typically neglected in the ISAC
literature~\cite{Ismail2024}.

Unlike~\cite{Costa2021}, where $Y_q^{(\nu)}(s,t;\thetain)$ is obtained from
the equivalent circuit of a specific unit-cell topology, we only assume
that, for fixed $t$, $\thetain$, and $\nu$, the meta-atom behaves as a
linear, causal, passive, and asymptotically stable one-port.
Hence, its admittance is positive-real~\cite{Anderson1973} and, by the bilinear
transformation in~\eqref{eq:meta-atom-reflection},
$\breve{\Gamma}^{(\nu)}_{\BP,q}(s,t;\thetain)$ is bounded-real, with all
its poles in the open left half-plane.
We further assume that:
\assumption[as:rational]{over the signal bandwidth, the reflection
coefficient is accurately described by a rational function of $s$ with
real coefficients and $\Np$ poles, without pole-zero cancellation;}
\assumption[as:simple]{the $\Np$ poles (either real or in complex 
conjugate pairs) are distinct.}

Denoting by $\Gamma_{\BP,q}^{(\nu)}$ the rational approximation of
$\breve{\Gamma}_{\BP,q}^{(\nu)}$ in~\textbf{(\ref{as:rational})}, the
response of the $q$-th meta-atom admits the following 
pole--residue representation:
\be
\Gamma_{\BP,q}^{(\nu)}(s,t;\thetain) =
\Gamma_{\infty,q}^{(\nu)}(t;\thetain) +
\sum_{p=1}^{\Np}
\frac{\zeta_{q,p}^{(\nu)}(t;\thetain)}{s-\rho_{q,p}^{(\nu)}(t;\thetain)},
\label{eq:pole-residue}
\ee
where $\rho_{q,p}^{(\nu)}(t;\thetain)$, with
$\Re\{\rho_{q,p}^{(\nu)}(t;\thetain)\}<0$, and
$\zeta_{q,p}^{(\nu)}(t;\thetain)$ are the poles and residues, and
$\Gamma_{\infty,q}^{(\nu)}(t;\thetain)$ is a real direct term accounting
for the non-resonant part of the response (for completeness, further details are provided in the supplementary material).
Unlike~\cite{Iudice2026}, where normal incidence and a single dominant
pole are assumed, \eqref{eq:pole-residue} may comprise multiple resonant
modes, whose poles and residues retain their dependence on $\thetain$ and
$\nu$.

From a signal-processing perspective, 
the function obtained by evaluating \eqref{eq:pole-residue} 
on the imaginary axis, that is,
\be
\Gamma^{(\nu)}_{\BP,q}(f,t;\thetain)\eqdef
\Gamma^{(\nu)}_{\BP,q}(s,t;\thetain)\big|_{s=j2\pi f},
\label{eq:pole-residue-f}
\ee
can be interpreted as the time-varying 
frequency response of the meta-atom
to an incident real \textit{bandpass} 
signal $x^{(\nu)}_{\BP,q}(t;\thetain)$ with center
frequency $f_c$.
Thus, using the standard representation of linear time-varying (LTV)
systems~\cite{Oppenheim1997Signals}, the complex-envelope representations
$x^{(\nu)}_{\BP,q}(t;\thetain)=\Re\{x^{(\nu)}_q(t;\thetain)\,e^{j2\pi f_ct}\}$
and $y^{(\nu)}_{\BP,q}(t;\thetain)=\Re\{y^{(\nu)}_q(t;\thetain)\,e^{j2\pi f_ct}\}$,
and the Hermitian symmetry of $\Gamma^{(\nu)}_{\BP,q}(f,t;\thetain)$ with
respect to $f$, which follows from \textbf{(\ref{as:rational})}, 
the \textit{baseband} input--output relation of the $q$-th
meta-atom reads as 
\be
y^{(\nu)}_q(t;\thetain)
=
\infint
\Gamma^{(\nu)}_q(f,t;\thetain)\, X^{(\nu)}_q(f;\thetain)\, e^{j2\pi ft}\, df,
\label{eq:reflected-lowpass-signal}
\ee
where $X^{(\nu)}_q(f;\thetain) \eqdef \mathcal{F}[x^{(\nu)}_q(t;\thetain)]$, and
$\Gamma^{(\nu)}_q(f,t;\thetain)\eqdef\Gamma^{(\nu)}_{\BP,q}(f+f_c,t;\thetain)$
is the time-varying \textit{baseband} frequency response of the meta-atom.
Incident fields comprising multiple plane waves or both polarizations can
be handled by superposition, since the meta-atom does not couple the two
polarizations; hence, in the following, we consider a single polarization
and omit the superscript $\nu$.


The input--output relation \eqref{eq:reflected-lowpass-signal} 
of the $q$-th meta-atom can be
equivalently expressed 
in the time domain~\cite{Bello1963} as
\be
y_q(t;\thetain)
=\infint
\gamma_q(\tau,t;\thetain)\, x_q(t-\tau;\thetain)\, d\tau,
\label{eq:reflected-lowpass-signal-time}
\ee
where the time-varying complex baseband impulse response
$\gamma_q(\tau,t;\thetain)$  is the inverse Fourier
transform of $\Gamma_q(f,t;\thetain)$ with respect to $f$, which, by
\eqref{eq:pole-residue-f}, is given by
\barr
\gamma_q(\tau,t;\thetain)
&= \Gamma_{\infty,q}(t;\thetain)\,\delta(\tau) \nonumber \\
&+ e^{-j2\pi f_c\tau}
\sum_{p=1}^{\Np}\zeta_{q,p}(t;\thetain)\,
e^{\rho_{q,p}(t;\thetain)\,\tau}\,\text{u}_{\infty}(\tau).
\label{eq:lowpass-meta-atom-impulse-response}
\earr
This causal response is the superposition of an instantaneous component
and $\Np$ exponentially decaying resonant modes, whose temporal extent is
governed by
\be
\rho_{\text{max}}
\eqdef
\max_{q\in\mathcal{J}_Q,\,p\in\mathcal{J}_{\Np},\,t\in\Rset,\,\thetain\in\Theta_{\text{s}}}
\Re\{\rho_{q,p}(t;\thetain)\}<0,
\label{eq:rho-max}
\ee
where $\Theta_{\text{s}}$ is the admissible range of AoA values.
For an attenuation level $0<\epsilon<1$, the \textit{effective memory} of
the STCM can be defined as
\be
\Tgamma\eqdef\frac{\ln(1/\epsilon)}{-\rho_{\text{max}}},
\label{eq:Tgamma}
\ee
beyond which every resonant mode is attenuated by at least a factor
$\epsilon$.
By the time--frequency scaling property of the Fourier
transform~\cite{Oppenheim1997Signals}, $\Tgamma$ also determines the
frequency scale of the meta-atom response, i.e.,
$B_\gamma\propto\Tgamma^{-1}$: a longer memory yields stronger frequency
selectivity over the signal bandwidth.

\subsection{Time modulation of the meta-atoms}
\label{sec:time-modulation}

The time-varying reflection response 
${\Gamma}_q(f,t; \thetain)$
is controlled by  
switching the state of 
each meta-atom~\cite{Darsena2024}.
For a varactor-loaded implementation, this can be achieved by applying a
{piecewise-constant} periodic control voltage:
\be
{v}_{\text{var},q}(t)
=
\text{rep}_{T}
\left[
\sum_{m=0}^{K-1}
v_{\text{var},q}^{[m]}
\, \text{u}_{\Tr}(t-m \Tr)
\right] \: ,
\label{eq:coded-gen}
\ee
where $T = K \Tr$  
is the modulation period 
and $v_{\text{var},q}^{[m]}$ 
is the constant voltage level applied during the 
$m$-th slot $[m \Tr, (m+1) \Tr[$ of duration $\Tr$.

Consistently with the adiabatic approximation recalled 
above, we assume that 
$
\Tgamma\ll \Tr
$.
Under this condition, switching transients can be neglected 
and each meta-atom reaches the steady state associated with the applied 
control value within each modulation slot. 
Accordingly, the circuit parameters can be modeled as piecewise 
constant over successive slots \cite{Minkov2017Exact, Ghasemi2021},
and the meta-atom frequency response inherits the same piecewise-constant 
periodic temporal structure:
\be
\Gamma_q(f, t;\thetain)
=\text{rep}_{T}
\left[
\sum_{m=0}^{K-1}
\Gamma_q^{[m]}(f;\thetain) \,
\text{u}_{\Tr}(t-m\Tr) \:
\right].
\label{eq:M-piecewise}
\ee
Here, $\Gamma_q^{[m]}(f;\thetain)$ is the steady-state baseband frequency
response of the meta-atom during the $m$-th slot, obtained by evaluating
\eqref{eq:pole-residue-f} at $f+f_c$, with poles
$\rho_{q,p}^{[m]}(\thetain)$, residues $\zeta_{q,p}^{[m]}(\thetain)$, and
direct term $\Gamma_{\infty,q}^{[m]}(\thetain)$ associated with the
control state applied during that slot.
Thus, each meta-atom is modeled as a frequency-selective 
spatially-dispersive LPTV
system \cite{Gardner1990} with period $T$ 
and piecewise-constant response in $t$.
%
\begin{table}[tb]
\centering
\caption{Hierarchy of meta-atom models.}
\label{tab:Gamma-models}
\footnotesize
\renewcommand{\arraystretch}{1.2}
\setlength{\tabcolsep}{3pt}
\begin{tabular}{p{1.5cm}p{1.5cm}p{3.0cm}p{1.8cm}}
\toprule
\textbf{Condition} & \textbf{Coding} & \textbf{Response} & \textbf{Model} \\
\midrule
$B_x\Tgamma\not\ll1$ & STCM & $\Gamma_q(f,t;\thetain)$ & FS, SD LPTV \\
$B_x\Tgamma\ll1$ & STCM
& $\bar{\Gamma}_q(t;\thetain)\eqdef\Gamma_q(0,t;\thetain)$ & FF, SD LPTV \\
$B_x\Tgamma\not\ll1$ & SCM & $\Gamma_q(f;\thetain)$ & FS, SD LTI \\
$B_x\Tgamma\ll1$ & SCM
& $\bar{\Gamma}_q(\thetain)\eqdef\Gamma_q(0;\thetain)$ & FF, SD LTI \\
$\thetain\approx\bar{\theta}_{\text{s}}$ & STCM/SCM
& $\Gamma_q(\cdot;\thetain)\approx\Gamma_q(\cdot;\bar{\theta}_{\text{s}})$
& SND \\
\bottomrule
\end{tabular}
\\[2pt]
\begin{minipage}{\columnwidth}
\scriptsize
FS: frequency-selective; FF: frequency-flat (memoryless);
SD: spatially dispersive; SND: spatially non-dispersive.
\end{minipage}
\end{table}
Consequently,  ${\Gamma}_{q}(f,t;\thetain)$ 
admits the Fourier-series expansion
\be
{\Gamma}_{q}(f,t;\thetain)
= \sum_{h\in\mathbb Z} b_{q}^{[h]}(f; \thetain) \, e^{j2\pi\frac{h}{T}t},
\label{eq:M-FS}
\ee
whose coefficients are
\be
b_{q}^{[h]}(f; \thetain)
= \frac{1}{K \Tr}
U_{\Tr}
\left(
\frac{h}{K \Tr}
\right)
\sum_{m=0}^{K-1}
\Gamma_q^{[m]}(f; \thetain)
\, 
e^{-j2\pi\frac{m h}{K}},
\label{eq:M-FS-coefficients}
\ee
where $U_{\Tr}(f) \eqdef \mathcal{F}[\text{u}_{\Tr}(t)] = \Tr \, \text{sinc}(f \Tr) \, e^{-j \pi f \Tr}$. 
%
The Fourier-series coefficients 
\eqref{eq:M-FS-coefficients} 
can also be expressed as
\begin{align}
b_{q}^{[h]}(f; \thetain)
&=
\underbrace{\frac{1}{K} \,
\sinc\left(\frac{h}{K}\right)
e^{-j\pi \frac{h}{K}} \, 
\bm{d}_h^\herm}_{=\bm{e}_h^\herm} \,
\bm{\Gamma}_{q}(f; \thetain) \nonumber \\
&=
\bm{e}_h^\herm \,
\bm{\Gamma}_{q}(f; \thetain) \: ,
\label{eq:bq_comp}
\end{align}
where
$\bm{d}_h
\eqdef
[1,e^{j2\pi h/K},\ldots,e^{j2\pi (K-1)h/K}]^\trasp \in \Cset^K$
and
$\bm{\Gamma}_{q}(f;\thetain)
\eqdef
[\Gamma_q^{[0]}(f;\thetain),\Gamma_q^{[1]}(f;\thetain),\ldots,
\Gamma_q^{[K-1]}(f;\thetain)]^\trasp \in \Cset^K$.
In~\eqref{eq:bq_comp}, the coding-dependent term
$\bm{d}_h^\herm \,\bm{\Gamma}_{q}(f;\thetain)$ is periodic in $h$ with period
$K$, i.e., $\bm{d}_{h+K}=\bm{d}_h$, and determines the amplitude and phase
with which each harmonic is excited; conversely, the nonperiodic envelope
$|\sinc(h/K)|$ governs the attenuation
of higher-order harmonics.

The above meta-atom model admits several limiting cases, depending on the effective
frequency support of $x_q(t;\thetain)$, denoted by
$\mathcal{W}_x \eqdef (f_{x,1},f_{x,2})$, with $f_{x,1}<0<f_{x,2}$, whose
bandwidth is $B_x \eqdef f_{x,2}-f_{x,1}$.
Since $B_\gamma \propto \Tgamma^{-1}$, the meta-atom response can be
regarded as \textit{frequency-flat} when $B_x \, \Tgamma \ll 1$, in which case
$\Gamma_q(f,t;\thetain) \approx \Gamma_q(0,t;\thetain) \eqdef
\bar{\Gamma}_q(t;\thetain)$ for all $f \in \mathcal{W}_x$.
Moreover, it becomes
\textit{time-invariant} when only space coding is implemented, as in a
\textit{space-coding metasurface} (SCM).
Finally, it can be treated as \textit{spatially non-dispersive} when the incident EM 
field is confined to a narrow angular
sector around a nominal AoA $\bar{\theta}_\text{s}$.
The resulting hierarchy of meta-atom models is summarized in
Table~\ref{tab:Gamma-models}.


\section{STCM-assisted ISAC system model}
\label{sec:system-model}

\begin{figure}[tb]
    \centering   
    \includegraphics[width=\columnwidth,trim=0 0 0 0,clip]{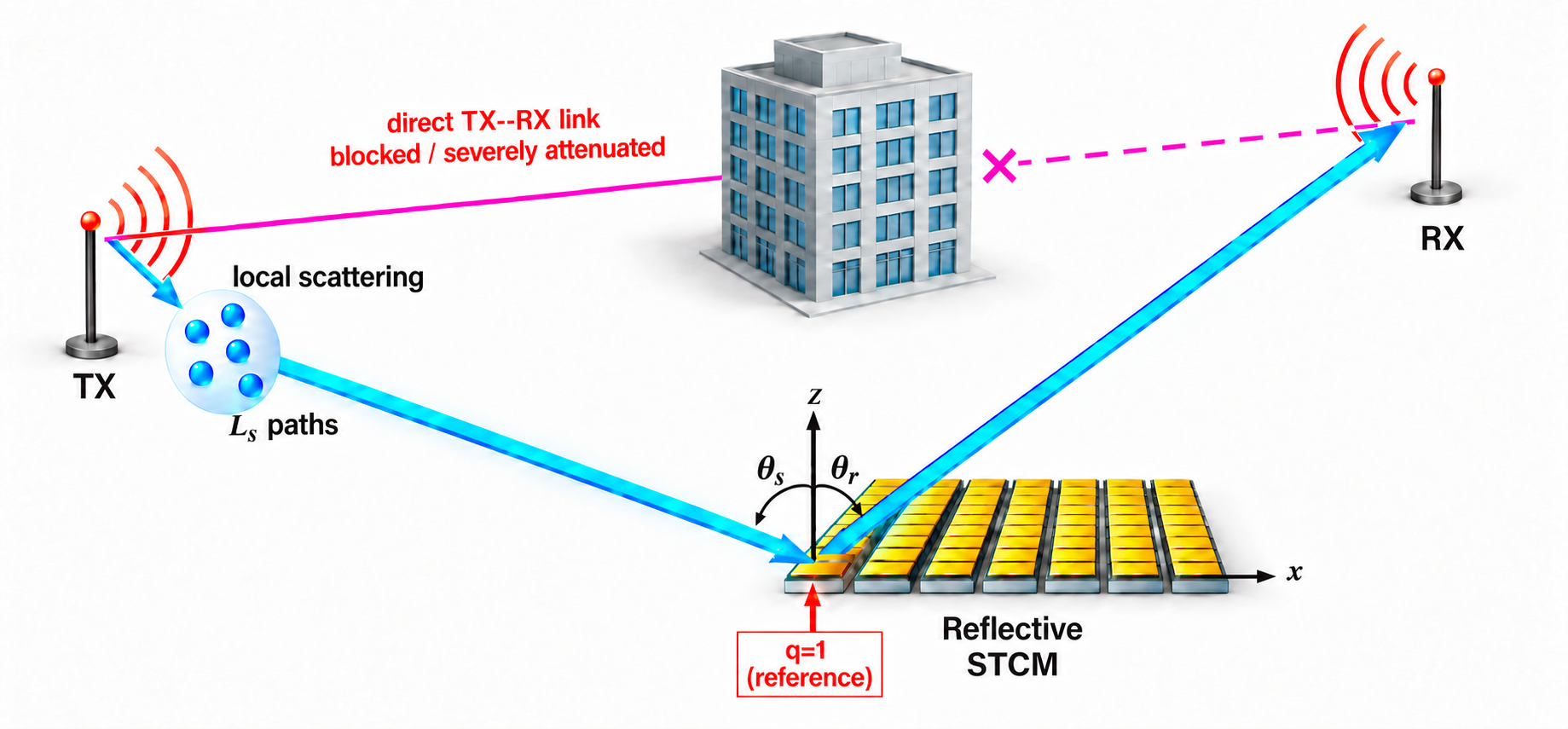}     
    \caption{The STCM-assisted ISAC system model.}
    \label{fig:ULA_RIS}
\end{figure}

As illustrated in Fig.~\ref{fig:ULA_RIS}, the considered STCM-assisted ISAC 
system comprises 
a single-antenna TX source,  a reflective STCM, and a single-antenna RX.%
\footnote{Extensions to multiple TX/RX pairs and multiantenna terminals 
are possible but are beyond the scope of the work.}
The TX emits an OFDM waveform $s(t)$, 
which is reflected by the STCM toward the RX. 
The direct TX--RX link is assumed to be
either blocked or severely attenuated, 
so that the received signal  
is dominated by the STCM-assisted 
propagation path.%
\footnote{If the direct link is non-negligible, it can be estimated 
separately by configuring the STCM to suppress
its reflected contribution and applying standard channel 
estimation methods \cite{Wang2020}.}
For tractability, we consider a two-dimensional geometry, in which the
TX, the RX, and all the propagation directions lie in the $(x,z)$ plane.
Since the STCM operates in 
reflection mode, both the TX and RX 
are assumed to lie in the half-plane $z>0$.
The AoAs and angles of departure (AoDs) are defined as 
signed angles measured from the positive $z$-axis, 
with positive angles evaluated in the 
counter-clockwise direction.
The position of the RX relative to the STCM is assumed to be known, 
so that both the AoD $\thetaout$ and the 
corresponding propagation delay $\tau_r$ 
are available. 
Conversely, the AoA $\thetain$ 
of the TX signal impinging on the STCM is unknown and must be estimated.

The STCM comprises $\Qc$ columns of meta-atoms,
uniformly spaced by a distance $d$ along the 
positive $x$-axis, with the first column serving as 
the spatial reference. 
All meta-atoms in the $q$-th column, 
with $q \in \mathcal{J}_{\Qc}$, 
are driven by a common temporal coding 
pattern~\cite{Cui2026} and therefore 
share the same time-varying reflection response,
which is modeled as described 
in Section~\ref{sec:meta-atom-model}. 
Column-wise control is also common in practical STCMs, since it simplifies the biasing circuitry.

In the following subsections, we present 
detailed models for the TX--STCM channel,  
the STCM processing, and the STCM--RX channel.
Throughout the analysis, both
the TX and the RX are assumed to lie in the far-field region of
the STCM.


\subsection{TX--STCM channel model}

The continuous-time complex-baseband OFDM signal 
transmitted by the TX source can be expressed as
\begin{align}
s(t) = \Ts^{-1/2}
\sum_{i=-\infty}^{\infty}\sum_{k=0}^{\Nc-1}
s_k[i]\, e^{j2\pi f_k t}\, \text{u}_{\Ts}(t-i\Ts),
\label{eq:TX-sig}
\end{align}
where $\Nc$ is the number of 
subcarriers, 
$f_k \eqdef (k-\lfloor \Nc/2 \rfloor) \Delta f$
denotes the $k$-th subcarrier frequency,
and $\Delta f = 1/\Tu$ is 
the subcarrier spacing (SCS), 
with $\Tu$ denoting the
useful OFDM symbol duration.%
\footnote{
The term $ \lfloor \Nc/2 \rfloor \Delta f$ in $f_k$ 
shifts the  subcarrier grid so that the passband 
OFDM spectrum is (approximately) centered
at the carrier frequency $f_c$.}
Accordingly, the nominal signal bandwidth is $\Bs \approx \Nc/\Tu$.
The corresponding discrete-time OFDM symbol 
comprises $\Nc$ samples spaced by $\Tc \eqdef \Tu/\Nc$. 
Moreover, $\Ts \eqdef \Tu+\Tcp$ 
is the total OFDM symbol duration, with 
$\Tcp$ denoting  the CP duration.
Finally, $s_k[i] \in \Cset$ in \eqref{eq:TX-sig}
denotes the symbol transmitted on the $k$-th subcarrier 
during the $i$-th OFDM symbol interval.
The symbols are modeled as a sequence of 
zero-mean, unit-variance complex random 
variables, with $s_{k_1}[i_1]$ and $s_{k_2}[i_2]$ 
statistically independent 
for $k_1 \ne k_2$ or $i_1 \ne i_2$.

The TX--STCM link is modeled as a \textit{frequency-selective block-fading} channel
with $\Ls$ resolvable paths, which remains constant 
over coherence
intervals of duration
$\Tcoh=\Lcoh\Ts$, comprising $\Lcoh\gg1$ OFDM symbols, 
and changes independently between consecutive
intervals.
Let $\tau_{\text{s},\ell,q}$ denote the propagation  
delay of the $\ell$-th path 
from the TX to the $q$-th column of STCM meta-atoms, 
for $\ell \in \mathcal{J}_{\Ls}$ 
and $q \in \mathcal{J}_{\Qc}$. 
For convenience, we define $\tau_{\text{s},\ell} \eqdef 
\tau_{\text{s},\ell,1}$ as the delay of the $\ell$-th 
path at the reference column ($q = 1$). 
Under the far-field assumption, 
the wavefront curvature across the
STCM aperture can be neglected, and the propagation delays at the
different columns are given by
\be
\tau_{\text{s},\ell,q}
=\tau_{\text{s},\ell}
+
(q-1) \tfrac{d}{c} 
\sin\theta_{\text{s},\ell}
\label{eq:delay}
\ee
where $\theta_{\text{s},\ell}$ denotes 
the AoA of the $\ell$-th path,  
$f_c$ and $\lambda_c=c/f_c$ are the carrier 
frequency and wavelength, 
respectively, with $c \approx 3\cdot 10^8$ m/s 
denoting the speed of light in vacuum.

For each $t$, 
let $\bm{h}_{\text{s}}(t)
\eqdef
[h_{\text{s},1}(t),h_{\text{s},2}(t),\ldots,h_{\text{s},\Qc}(t)]^{\trasp}
\in\Cset^{\Qc\times 1}$
denote the vector collecting the complex-baseband channel 
impulse responses (CIRs) from the TX to 
the $\Qc$ STCM columns. 
Its $q$-th entry is
\be
h_{\text{s},q}(t)
= \sum_{\ell=1}^{L_{\text{s}}}
\alpha_{\text{s},\ell}\,
e^{-j2\pi f_c\tau_{\text{s},\ell,q}}\,
\delta(t-\tau_{\text{s},\ell,q}) \: .
\label{eq:h-def}
\ee
We further adopt the \textit{local-scattering model}
of \cite{Bjornson2017}, according to which the dominant
scatterers form a compact cluster
in the vicinity of the TX (see Fig.~\ref{fig:ULA_RIS}).
If the spatial extent of this cluster is much smaller
than the TX--STCM distance, its angular spread,
as observed from the STCM, is negligible.
Consequently, the path-dependent AoAs
$\theta_{\text{s},\ell}$ can be approximated by
a common but unknown nominal angle $\theta_{\text{s}}$,
Hence, by setting 
$\theta_{\text{s},\ell}\approx\theta_{\text{s}}$,
\eqref{eq:delay} 
can be rewritten as
\be
\tau_{\text{s},\ell,q}
\approx \tau_{\text{s},\ell}
+
(q-1) \tfrac{\psi_{\text{s}}}{f_c}
\label{eq:delay-approx}
\ee
where $\psi_{\text{s}} \eqdef
\lambda_c^{-1} \, d \, \sin\theta_{\text{s},\ell}$
is the normalized spatial 
frequency  associated with 
the AoA $\theta_\text{s}$~\cite{Ma2021}.%
\footnote{Should the local-scattering assumption fail,
so that the angular spread is non-negligible, the meta-atom model of
Section~\ref{sec:meta-atom-model} remains valid by superposition, at the cost of
an angular estimation problem involving multiple unknown AoAs, which is
beyond the scope of this work.}
The propagation delays $\tau_{\text{s},\ell}$, however,
remain distinct in \eqref{eq:delay-approx}, since the multipath components are assumed to be
resolvable in the delay domain, i.e.,
$\left|\tau_{\text{s},\ell}-\tau_{\text{s},\ell'}\right|\gtrsim \Bs^{-1}$
for $\ell\neq\ell'$.
Note that a negligible angular spread implies a common AoA $\thetain$ for all
incident multipath components, but not an angle-independent meta-atom
response. Since this common AoA is unknown and ranges over
$\Theta_{\text{s}}$, its effect must be retained explicitly. The
spatially nondispersive model in the last row of
Table~\ref{tab:Gamma-models} would instead require the meta-atom
response to be approximately invariant 
over the entire angular search
set.

Using approximation~\eqref{eq:delay-approx}, 
the CIR in \eqref{eq:h-def}
reduces to
\be
h_{\text{s},q}(t)
\approx e^{-j2\pi(q-1)\psi_{\text{s}}}
\sum_{\ell=1}^{L_{\text{s}}}
g_{\text{s},\ell} \,
\delta(t-\tau_{\text{s},\ell,q}) 
\label{eq:h-local-scattering}
\ee
where
$g_{\text{s},\ell}\triangleq\alpha_{\text{s},\ell}\,e^{-j2\pi f_c\tau_{\text{s},\ell}}$
is the effective complex gain of the $\ell$-th path.
Consequently, the complex-baseband signal incident 
on the meta-atoms of the $q$-th STCM column 
is
\barr
x_q(t; \thetain)
= 
e^{-j2\pi(q-1)\psi_{\text{s}}}
\sum_{\ell=1}^{L_{\text{s}}}
g_{\text{s},\ell} \,
s(t-\tau_{\text{s},\ell,q}), 
\label{eq:x-RIS}
\earr
with
$s(t)$ given by \eqref{eq:TX-sig}.
The Fourier transform of \eqref{eq:x-RIS} is
$
X_q(f; \thetain)
=
e^{-j2\pi(q-1)\psi_{\text{s}}} \, G_{\text{s},q}(f) \, S(f)
$,
where
\begin{multline}
S(f)
=
\Ts^{1/2}
\sum_{i=-\infty}^{\infty}
\sum_{k=0}^{\Nc-1}
s_k[i] \,
\text{sinc}
\left[
\left( f-f_k \right) \Ts
\right] \\
\times
e^{-j2\pi\left(f-f_k\right)
\left(\frac{\Ts}{2}+i \Ts\right)}
\label{eq:TX-sig-freq}
\end{multline}
is the Fourier transform of \eqref{eq:TX-sig} and
$G_{\text{s},q}(f) \eqdef \mathcal{F}[g_{\text{s},q}(t)]$,
with ${g}_{\text{s},q}(t) \eqdef\sum_{\ell=1}^{\Ls}g_{\text{s},\ell}\,
\delta(t-\tau_{\text{s},\ell,q})$.

Since the TX--STCM channel is assumed to be LTI within each coherence
interval, it does not expand the spectral support of the OFDM
signal. Therefore, the signal impinging on each meta-atom
retains the bandwidth of the OFDM TX signal, i.e., $B_x=B_s$.


\subsection{STCM processing}
\label{sec:STCM-processing}

For the STCM input-output processing, we build upon the
physics-based harmonic representation developed in
Section~\ref{sec:meta-atom-model}.
We set the STCM modulation period equal to the useful OFDM
symbol duration, 
so that the harmonic spacing $1/T$ coincides with the OFDM
SCS $1/\Tu$ \cite{Iudice2026}.
As a result, the harmonic components 
generated by the temporal modulation
are aligned with the OFDM frequency grid.
Accordingly, 
by using 
\eqref{eq:bq_comp},
\eqref{eq:M-FS} can be rewritten for $T = \Tu$ as
\barr
{\Gamma}_{q}(f,t;\thetain)
&=
\sum_{h\in\mathbb Z}
\bm{e}_h^\herm \, 
\boldsymbol{\Gamma}_{q}(f;\thetain) \, 
e^{j2\pi\frac{h}{\Tu}t}  \nonumber \\
&= \frac{1}{K}
\sum_{h\in\mathbb Z}
\sinc\left(\frac{h}{K}\right)
e^{-j\pi \frac{h}{K}}
\bm{d}_h^\herm \, 
\bm{\Gamma}_{q}(f;\thetain) \, 
e^{j2\pi\frac{h}{\Tu}t} \: .
\label{eq:M-FS-vector}
\earr
As discussed in Section~\ref{sec:time-modulation}, this representation
assumes the adiabatic condition $\Tgamma\ll\Tr$, which should not be
confused with frequency flatness, governed instead by
$B_x\Tgamma$ (see Table~\ref{tab:Gamma-models}).
Indeed, since $\Tr=\Tu/K$ and $B_x\approx\Nc/\Tu$, the adiabatic
condition becomes $K\ll\Nc/(B_x\Tgamma)$. This condition, in the
frequency-selective regime, imposes an increasingly restrictive
upper bound on $K$ as $B_x\Tgamma$ grows, whereas it is generally less
restrictive in the frequency-flat regime.


\subsection{STCM--RX channel model}

The STCM is deployed so as to ensure an unobstructed propagation path
toward the RX, whose position relative to the STCM is known and which lies in
the far field of the STCM.
Consequently, the STCM--RX channel is dominated by its LoS
component~\cite{Ozdogan2020}, with negligible multipath delay spread, and
the field reflected toward the RX can be approximated as a plane wave with known AoD
$\thetaout$.
The CIR between the $q$-th STCM column and the RX is
\be
h_{\text{r},q}(t) =
g_{\text{r}}\,
e^{-j2\pi(q-1)\psi_{\text{r}}}\,
\delta(t-\tau_{\text{r},q}),
\qquad q\in\mathcal{J}_{\Qc},
\label{eq:STCM-RX-channel}
\ee
where $\tau_{\text{r},q}\eqdef\tau_{\text{r}}+(q-1)\psi_{\text{r}}/f_c$,
with $\tau_{\text{r}}$ denoting the propagation delay from the reference STCM
column, $g_{\text{r}}\in\Cset$ the corresponding complex gain, including
the carrier phase $e^{-j2\pi f_c\tau_{\text{r}}}$, and
$\psi_{\text{r}}\eqdef\lambda_c^{-1}d\sin\thetaout$.
Since the RX position is known, $\tau_{\text{r}}$ and $\thetaout$ are
treated as known geometric parameters.
Apart from the pure delay $\tau_{\text{r},q}$, the channel seen by each
column is thus \textit{frequency-flat}, whereas the differential delays
$(q-1)\psi_{\text{r}}/f_c$ give rise to a frequency-dependent spatial phase
profile; the STCM--RX channel is also assumed constant over the coherence
interval.

Note that, from \eqref{eq:reflected-lowpass-signal} and \eqref{eq:M-FS}
with $T=\Tu$, the spectrum reflected by the $q$-th STCM column is
$Y_q(f;\thetain)=\sum_{h\in\Zset}b_q^{[h]}(f-h/\Tu;\thetain)\,X_q(f-h/\Tu;\thetain)$,
i.e., a superposition of harmonically shifted copies of the incident
spectrum, whose amplitudes decay according to the sinc envelope
in~\eqref{eq:bq_comp}; 
the STCM--RX channel is thus required to be
frequency-flat only over the effective bandwidth $B_y$, defined by retaining the harmonics
whose power exceeds a prescribed threshold.

Using \eqref{eq:STCM-RX-channel}, the complex-baseband signal at the RX is
\be
r(t;\thetain)
=
g_{\text{r}}
\sum_{q=1}^{\Qc} e^{-j2\pi(q-1)\psi_{\text{r}}}\,
y_q(t-\tau_{\text{r},q};\thetain) + v(t),
\label{eq:rr-signal-1}
\ee
where $v(t)$ is zero-mean additive white Gaussian noise 
(AWGN) with autocorrelation
$R_{vv}(\tau)\eqdef\Es[v(t)v^*(t-\tau)]=2N_0\,\delta(\tau)$.
Owing to the causality and finite effective memory of
$\gamma_q(\tau,t;\thetain)$, substituting
\eqref{eq:reflected-lowpass-signal-time} and \eqref{eq:x-RIS} into \eqref{eq:rr-signal-1} yields
\begin{multline}
r(t;\thetain) \approx
g_{\text{r}}
\sum_{q=1}^{\Qc}
\sum_{\ell=1}^{\Ls}
g_{\text{s},\ell}\,
e^{-j2\pi(q-1)(\psi_{\text{s}}+\psi_{\text{r}})} \\
\times \int_{0}^{\Tgamma} \!\!\!
\gamma_q(\tau,t-\tau_{\text{r},q};\thetain)\,
s(t-\tau_{\text{s},\ell,q}-\tau_{\text{r},q}-\tau)\, d\tau + v(t).
\label{eq:rr-signal-dispersive}
\end{multline}


\subsection{Harmonic extraction}
\label{subsec:ofdm-harmonic}

The RX is assumed to achieve perfect 
timing synchronization with the earliest 
arriving signal component over the cascaded TX–STCM–RX channel. 
Such synchronization is 
assumed to be performed directly 
from the received waveform and does not require prior 
knowledge of the unknown AoA $\thetain$.
Moreover, for notational 
convenience, the common known propagation
delay $\tau_{\text{r}}$ of the STCM--RX link
is absorbed into the timing reference and
henceforth set to zero.

Let
$
\tau_{0} \eqdef
\min_{\ell\in \mathcal{J}_{\Ls},\,q\in\mathcal{J}_{\Qc}}
\left\{
\tau_{\text{s},\ell,q}
+
\tau_{\text{r},q}
\right\}
$
denote the earliest propagation 
delay of the cascaded channel.
The $h$-th harmonic component ($h\in\mathbb Z$)
is extracted through conventional OFDM demodulation.%
\footnote{For notational convenience, OFDM receiver processing
is represented in continuous time, whereas practical receivers
perform subcarrier demodulation in the digital domain using
fast Fourier transform (FFT) algorithms.}
Specifically, after discarding the CP, the 
received signal is downconverted by $f_h$ 
and integrated over the useful 
interval of the $n$-th OFDM symbol, yielding
\be
r_h[n;\thetain]
\eqdef
\Ts^{-1/2}
\int_{\tau_{0} + n\Ts+\Tcp}^{\tau_{0}+ (n+1)\Ts}
r(t;\thetain) \, e^{-j2\pi f_h t}\,dt \,.
\label{eq:yh_ofdm_def}
\ee
Let
$\Delta\tau_{\text{s,max}}
\eqdef
\max_{\ell\in\mathcal J_{\Ls}}\tau_{\text{s},\ell}
-
\min_{\ell\in\mathcal J_{\Ls}}\tau_{\text{s},\ell}$
denote the delay spread of the TX--STCM channel.
Under the local-scattering assumption, the overall
effective delay spread of the cascaded channel,
including the meta-atom memory, is
$
T_{\text{eff}}
=
\Delta\tau_{\text{s,max}}
+
\Tgamma
+
(\Qc-1)\tfrac{d}{c}
\left|
\sin\thetain+\sin\thetaout
\right|.
$
Since
$\left|\sin\thetain+\sin\thetaout\right|\le 2$,
a sufficient condition for the CP to accommodate
the overall channel memory, irrespective of the
unknown AoA, is
\be
\Tcp
\geq
\Delta\tau_{\text{s,max}}
+
\Tgamma
+
2(\Qc-1) \tfrac{d}{c}.
\label{eq:CP-cond}
\ee
Under this condition, the integration interval
in \eqref{eq:yh_ofdm_def} is free of
intersymbol interference (ISI)~\cite{Iudice2026},
so that only the $n$-th transmitted OFDM symbol
contributes to $r_h[n;\thetain]$.

Substituting \eqref{eq:rr-signal-dispersive} 
into \eqref{eq:yh_ofdm_def},
plugging \eqref{eq:M-FS-vector} and 
\eqref{eq:TX-sig} into the result, 
and exploiting subcarrier orthogonality 
over the useful OFDM interval $\Tu$,
yields
the expression reported in \eqref{eq:yh_ofdm_scalar},
where $\ku \eqdef {\Tu}/{\Ts} \in (0,1)$,
$G_{s}(f) \eqdef G_{s,1}(f)$, 
and $v_{h}[n]$ is the demodulated
noise term, defined similarly 
to \eqref{eq:yh_ofdm_def}, having zero mean and
variance $\sigma_v^2 \eqdef 2 N_0 \ku $.
%
%
%
\begin{figure*}[!t]
\normalsize
\setcounter{temp_counter}{\value{equation}}
%
\barr
r_{h}[n;\thetain]
&=
\ku \, g_{\text{r}} 
\sum_{k=0}^{\Nc-1}
s_k[n]
\sum_{\ell=1}^{\Ls}
g_{\text{s},\ell}\,
e^{-j 2 \pi f_k \tau_{\text{s},\ell}} \,
\bm{e}_{h-k}^\herm
\sum_{q=1}^{\Qc}
\bm{\Gamma}_q\left(f_k; \thetain \right)
e^{-j2\pi(q-1)\left(1+\frac{f_k}{f_c}\right)\psi_{\text{s}}}\,
e^{-j2\pi(q-1)\left(1+\frac{f_h}{f_c}\right)\psi_{\text{r}}}
+
v_{h}[n] \nonumber
\\
&= \ku \, g_{\text{r}} 
\sum_{k=0}^{\Nc-1}
s_k[n] \,
G_{\text{s}}\left( f_k \right)
\bm{e}_{h-k}^\herm \,
\bm{R}(f_k; \thetain)\, \bm{A}_\text{r}(f_h) \,
\bm{a}_\text{s}(f_k;\thetain)
+
v_{h}[n],
\label{eq:yh_ofdm_scalar}
\earr
\setcounter{equation}{\value{temp_counter}}
\hrulefill
\vspace*{4pt}
\end{figure*}
\addtocounter{equation}{1}
%
Moreover, the STCM coding-pattern matrix
$\bm{R}(f_k; \thetain) \in\Cset^{K\times \Qc}$
in \eqref{eq:yh_ofdm_scalar}, given by
\be
\bm{R}(f_k; \thetain)
\eqdef
\left[
\bm{\Gamma}_1(f_k; \thetain),
\bm{\Gamma}_2(f_k; \thetain),
\ldots,
\bm{\Gamma}_{\Qc}(f_k; \thetain)
\right],
\ee
collects the space--time reflection coefficients of the $\Qc$
meta-atom columns at the $k$th OFDM subcarrier, with its $(m,q)$-th
entry corresponding to the response of the $q$-th column during the
$m$-th modulation slot.
Finally,
\be
\bm{a}_{\text{s}}(f;\thetain)
\eqdef 
\left[
1,
e^{-j2\pi\left(1+\frac{f}{f_c}\right)\psi_{\text{s}}},
\ldots,
e^{-j2\pi(\Qc-1)\left(1+\frac{f}{f_c}\right)\psi_{\text{s}}}
\right]^\trasp
\ee
is the $\Qc$-dimensional steering vector 
associated with the unknown TX--STCM AoA $\thetain$,
whereas the known spatial phase profile
associated with the STCM--RX link
is represented by the $\Qc \times \Qc$ 
diagonal steering matrix
\be
\bm{A}_{\text{r}}(f)
\eqdef
\diag
\left(
1,
e^{-j2\pi\left(1+\frac{f}{f_c}\right)\psi_{\text{r}}},
\ldots,
e^{-j2\pi(\Qc-1)\left(1+\frac{f}{f_c}\right)\psi_{\text{r}}}
\right).
\ee
It is worth noting that both $\bm{a}_{\text{s}}(f;\thetain)$ and
$\bm{A}_{\text{r}}(f)$ depend on frequency
through the \textit{effective wavelength} $\lambda_f \eqdef c/(f_c+f)$, since
$(1+f/f_c)\,\psi =\lambda_f^{-1}d\sin\theta$. This is the well-known
spatial-wideband or \textit{beam-squint} effect~\cite{Ma2021}, which arises
because the differential propagation delays across the STCM aperture are
retained in the complex envelopes, rather than only in the carrier phases.
In the narrowband-array limit $(\Qc-1)\,d/c \ll B_s^{-1}$, one has
$\bm{a}_{\text{s}}(f_k;\thetain)\approx\bm{a}_{\text{s}}(0;\thetain)$
and $\bm{A}_{\text{r}}(f_h)\approx\bm{A}_{\text{r}}(0)$, and
\eqref{eq:yh_ofdm_scalar} reduces to the model commonly adopted in the
STCM literature.

\begin{remark}[Structured intercarrier coupling induced by the STCM]
\label{rem:structured-ICI}
Equation~\eqref{eq:yh_ofdm_scalar} shows that, when the STCM operates
in time-varying mode, the signal observed in the $h$-th OFDM bin
generally contains contributions from all transmitted subcarriers
($k \in \mathcal{O}_{\Nc}$). 
Specifically, the contribution of the $k$-th subcarrier is
transferred to the $h$-th bin through the STCM harmonic of order
$h-k$.
The choice $T=\Tu$ aligns the harmonic spacing $1/T$ with the OFDM
subcarrier spacing $1/\Tu$. Hence, these frequency-shifted components
fall exactly on the OFDM grid and can be extracted by conventional
CP removal and FFT processing, without requiring  a dedicated 
sensing receiver or additional RF hardware. This alignment prevents off-grid
spectral leakage but does not remove the coupling among
subcarriers, which takes the form of 
 a \textit{structured} form of intercarrier interference
(ICI), whose coefficients are determined by the known STCM coding
pattern and by the meta-atom responses. During acquisition, this
controlled coupling is deliberately exploited to obtain multiple
harmonic-domain observations for AoA and channel estimation.
When temporal modulation is disabled during data transmission, only
the zero-order harmonic remains. Consequently, the terms with
$h\neq k$ vanish, \eqref{eq:yh_ofdm_scalar} reduces to a diagonal
per-subcarrier OFDM model, and no STCM-induced ICI is present.
\end{remark}


\section{Dual-mode STCM operating protocol}
\label{sec:dual-mode-protocol}

\begin{figure}[t]
\centering
\includegraphics[width=0.98\columnwidth]{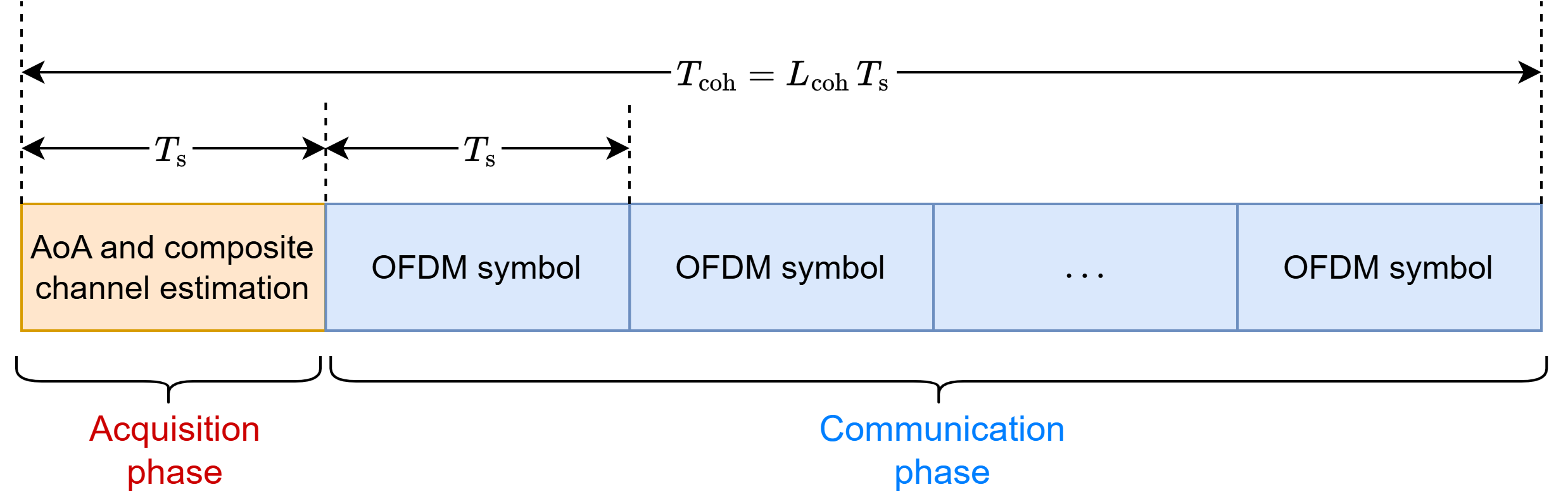}
\caption{Time structure of the dual-mode STCM protocol
operating over a channel coherence interval of $L_{\text{coh}}$ OFDM symbols
each of length $\Ts$.}
\label{fig:phase}
\end{figure}

The proposed dual-mode protocol operates according 
to the block-fading model of the TX--STCM channel,
as shown in Fig.~\ref{fig:phase}. 
Each coherence interval of duration $\Tcoh=\Lcoh \, \Ts$ 
comprises $\Lcoh \gg 1$ OFDM symbols, and
is divided into an \textit{acquisition phase} 
devoted to sensing and channel
estimation, followed by a \textit{communication phase}.

During the acquisition phase, the STCM operates in time-varying mode,
and the TX transmits a single dedicated pilot OFDM symbol, whose
STCM-generated harmonics are processed at the RX to estimate the AoA
$\thetain$ and the composite channel.
To this end, the TX, the RX, and the STCM controller are assumed to be
synchronized at the OFDM-symbol level, so that the pilot symbol coincides
with the activation of the time-modulation pattern, which is in turn
aligned with the RX observation window.
Before the communication phase, the STCM is switched to the static
configuration computed from the estimated AoA, whereas the estimated
composite channel is used for OFDM frequency-domain equalization at the
RX.

The acquisition procedure is repeated at the beginning of each
coherence block to track variations in the TX--STCM propagation
environment and update both the STCM configuration and the
composite channel estimate. 
The additional DoFs 
provided by  temporal coding at the STCM are thus
exploited only during acquisition, 
while conventional static
beamforming is employed during data transmission.
The following subsections develop the acquisition and
communication procedures underlying the proposed protocol.


\subsection{VC-aided joint AoA and channel estimation}
\label{subsec:vc-aided-acquisition}

During the acquisition phase, the AoA and the composite channel are
jointly estimated from a single pilot OFDM symbol, in which only $P<\Nc$
uniformly spaced subcarriers, with indexes $k_p=k_0+pD$,
$p\in\mathcal{O}_P$, are activated as pilots, with $k_0$ and $D$ being the
first pilot index and the pilot spacing, respectively; the remaining
$\Nc-P$ subcarriers are configured as VCs, i.e., 
they are intentionally left unmodulated.

Without loss of generality, we choose the time reference origin such
that the pilot OFDM symbol transmitted during the acquisition phase
has index $n=0$.
Since only this OFDM symbol is considered throughout the acquisition
procedure, we omit its time index and write
$r_h \eqdef r_h[0;\thetain]$, $v_h\eqdef v_h[0]$, and
$s_k\eqdef s_k[0]$, 
where the dependence on $\thetain$ in $r_h$ is also
dropped. 
Specializing \eqref{eq:yh_ofdm_scalar} to the pilot subcarriers
$k=k_p$, $p\in\mathcal{O}_P$, which are the only ones with $s_k\neq0$,
yields
\begin{align}
r_h
&=
\sum_{p=0}^{P-1}
\beta_{k_p}s_{k_p}
\bm{e}_{h-k_p}^{\herm}
\bm{R}(f_{k_p}; \thetain)\,
\bm{A}_{\text{r}}(f_h) \,
\bm{a}_{\text{s}}(f_{k_p};\thetain)
+v_h  \nonumber
\\
&=
\frac{1}{K}
\sum_{p=0}^{P-1}
\beta_{k_p}s_{k_p}
\sinc\left(\frac{h-k_p}{K}\right)
e^{-j\pi\frac{h-k_p}{K}}
\label{eq:r-pilot-sinc}
\\
&\quad\times
\bm{d}_{h-k_p}^{\herm}
\bm{R}(f_{k_p}; \thetain)\,
\bm{A}_{\text{r}}(f_h) \,
\bm{a}_{\text{s}}(f_{k_p};\thetain)
+v_h , \nonumber
\end{align}
where the second equality
follows from 
\eqref{eq:bq_comp},
and $\beta_{k} \eqdef \ku\,g_{\text{r}}\, G_{\text{s}}(f_{k}) \in \Cset$
denotes 
the composite channel coefficient associated with the
$k$-th subcarrier, 
which accounts for the frequency-selective
TX--STCM channel and the common gain of the STCM--RX link, while
excluding the reflection response and spatial combining introduced
by the STCM. 

For each pilot subcarrier $k_p$, we consider only the $2H+1$
STCM-induced harmonic components falling on the OFDM bins
$\mathcal{M}_p\eqdef\{k_p-H,\ldots,k_p,\ldots,k_p+H\}$,
with $1\leq H\leq \lfloor (K-1)/2\rfloor$.
According to~\eqref{eq:bq_comp}, harmonics whose indexes differ by $K$
share the same coding-dependent term and differ only in the sinc
envelope, which decreases with the harmonic order.
Hence, since $2H+1\le K$, each coding-dependent term is retained at most
once, through its strongest replica, i.e., the one closest to the pilot
subcarrier.
The harmonics produced by STCM modulation 
outside this range are not identically zero, but
their amplitudes are attenuated by the sinc envelope.

The first pilot index $k_0$ and the pilot spacing $D$ are selected
such that all the retained bins fall within the OFDM band, i.e.,
\be
k_0\geq H,
\qquad
k_0+(P-1)D+H\leq \Nc-1 .
\label{eq:k0-harmonic-condition}
\ee
By setting $k_0=H$, \eqref{eq:k0-harmonic-condition} is satisfied for
any $D\leq(\Nc-1-2H)/(P-1)$, with $P>1$.
Accordingly, we choose the largest admissible pilot spacing as
\be
\bar{D} \eqdef\left\lfloor\frac{\Nc-1-2H}{P-1}\right\rfloor ,
\label{eq:pilot-spacing}
\ee
which satisfies $\bar{D} \geq 1$ provided that $P\leq \Nc-2H$.
Thus, under~\eqref{eq:k0-harmonic-condition}, $\mathcal{M}_p\subseteq\mathcal{O}_{\Nc}$
for all $p\in\mathcal{O}_P$, and the overall set of OFDM bins
considered for acquisition is
\be
\mathcal{M}
\eqdef
\bigcup_{p=0}^{P-1}\mathcal{M}_p
\subseteq \mathcal{O}_{\Nc}.
\label{eq:overall-harmonic-set}
\ee
By arranging the elements of $\mathcal{M}$ in increasing order, this
set can be written as
$
\mathcal{M}
=
\{\mu_0,\mu_1,\ldots,\mu_{M-1}\}$,
with
$\mu_0<\mu_1<\cdots<\mu_{M-1}$, whose cardinality
is
$
M = 2H+1+(P-1)\min\{\bar{D},2H+1\}$.
Thus, since $H\geq 1$, it follows that
\be
P
<
P+2H
\leq
M
\leq
P(2H+1)\: ,
\label{eq:M-bounds}
\ee
where the lower bound $M=P+2H$ is attained for $\bar{D}=1$, 
corresponding to the maximum overlap 
among the pilot-centered sets, whereas the
upper bound $M=P(2H+1)$ is attained when 
these sets are mutually disjoint, i.e., 
for $\bar{D}\geq 2H+1$ or, equivalently,
by exploiting~\eqref{eq:pilot-spacing}, for
$P\leq \left\lfloor \Nc/(2H+1) \right\rfloor$.

For the $p$-th pilot subcarrier, we introduce the
harmonic-projection matrix
$\bm{E}_{k_p}\eqdef
[\bm{e}_{\mu_0-k_p},\bm{e}_{\mu_1-k_p},\ldots,\bm{e}_{\mu_{M-1}-k_p}]^{\herm}
\in\Cset^{M\times K}$
and the corresponding effective response vector
\be
\widetilde{\bm{b}}_{k_p}(\thetain)
\eqdef
\big[\bm{E}_{k_p}\,\bm{R}(f_{k_p}; \thetain)\odot\widetilde{\bm{A}}_{\text{r}}\big]\,
\bm{a}_{\text{s}}(f_{k_p};\thetain)
\in\Cset^{M},
\label{eq:btilde-pilot-M}
\ee
where the $m$-th row of the known matrix
$\widetilde{\bm{A}}_{\text{r}}\in\Cset^{M\times \Qc}$ collects the diagonal
entries of $\bm{A}_{\text{r}}(f_{\mu_m})$, i.e.,
$\{\widetilde{\bm{A}}_{\text{r}}\}_{m,q}
=e^{-j2\pi(q-1)\left(1+\frac{f_{\mu_m}}{f_c}\right)\psi_{\text{r}}}$.
Stacking \eqref{eq:r-pilot-sinc} over all the OFDM bins
$\mu_m\in\mathcal{M}$
then yields the acquisition model
\be
\bm{r}
=
\sum_{p=0}^{P-1}
\beta_{k_p}s_{k_p}
\widetilde{\bm{b}}_{k_p}(\thetain)
+
\bm{v} \: ,
\label{eq:r-acquisition-sum}
\ee
where
$\bm{r}
\eqdef
[r_{\mu_0},
r_{\mu_1},
\ldots,
r_{\mu_{M-1}}]^\trasp
\in\Cset^{M}$
and
$\bm{v}
\eqdef
[v_{\mu_0},
v_{\mu_1},
\ldots,
v_{\mu_{M-1}}]^\trasp
\in\Cset^{M}$.
Model~\eqref{eq:r-acquisition-sum} can be interpreted
as a single-snapshot observation of an $M$-element virtual array, whose
elements are the retained harmonic bins in $\mathcal{M}$: each pilot
contributes through a \textit{virtual steering vector}
$\widetilde{\bm{b}}_{k_p}(\thetain)$, associated with the same unknown
direction $\thetain$ and weighted by the known pilot symbol $s_{k_p}$ and
the unknown channel coefficient $\beta_{k_p}$.
Unlike a conventional physical array, the virtual steering vector depends
on the pilot subcarrier, owing to the harmonic shift introduced by
$\bm{E}_{k_p}$, the frequency selectivity of $\bm{R}(f_{k_p};\thetain)$, and
beam squint, and can be shaped through the STCM coding pattern.

Equation~\eqref{eq:r-acquisition-sum} can be rewritten 
more compactly as
\be
\bm{r}
=
\widetilde{\bm{B}}_{\pil}(\thetain) \, 
\bm{S}_{\pil} \, 
\boldsymbol{\beta}_{\pil}
+
\bm{v} \: ,
\label{eq:compact-model-gen}
\ee
where
$\bm{S}_{\pil} \eqdef
\text{diag}
\left(
s_{k_0},
s_{k_1},
\ldots,
s_{k_{P-1}}
\right)
\in\Cset^{P\times P}
$
is the known diagonal pilot matrix,
$
\widetilde{\bm{B}}_{\pil}(\thetain)
\eqdef
[
\widetilde{\bm{b}}_{k_0}(\thetain),
\widetilde{\bm{b}}_{k_1}(\thetain),
\ldots,
\widetilde{\bm{b}}_{k_{P-1}}(\thetain)
]
\in\Cset^{M\times P}
$,
and 
$
\boldsymbol{\beta}_{\pil}
\eqdef
[
\beta_{k_0},
\beta_{k_1},
\ldots,
\beta_{k_{P-1}}
]^\trasp
\in\Cset^P
$
collects the unknown composite 
channel coefficients associated with the
pilot subcarriers.

The objective of the acquisition phase
is to jointly estimate
the AoA $\thetain$ and the composite
channel vector $\boldsymbol{\beta}_{\pil}$
from the observation model 
in \eqref{eq:compact-model-gen}.
Since $\bm{v} \sim \mathcal{CN}(\bm{0},\sigma_v^2 \, \bm{I}_{M})$,
their joint ML estimate is given by
\be
(
\widehat{\theta}_{\text{s}},
\widehat{\boldsymbol{\beta}}_{\pil}
)
=
\arg\min_{\substack{
\thetain\in\Theta_\text s\\
\boldsymbol{\beta}_{\pil}\in\Cset^P
}}
\left\|
\bm{r}
-
\widetilde{\bm{B}}_{\pil}(\thetain) \,
\bm{S}_{\pil} \,
\boldsymbol{\beta}_{\pil}
\right\|^2 \: .
\label{eq:joint-ml-theta-beta}
\ee
For a fixed value of $\thetain$, minimization with
respect to $\boldsymbol{\beta}_{\pil}$ reduces to 
a standard \textit{least-squares} (LS) 
problem, whose solution \cite{Kay1993} is
\be
\widehat{\boldsymbol{\beta}}_{\pil}(\thetain)
=
\left[
\bm{S}_{\pil}^{\herm} \, 
\widetilde{\bm{B}}_{\pil}^{\herm}(\thetain) \, 
\widetilde{\bm{B}}_{\pil}(\thetain) \, 
\bm{S}_{\pil}
\right]^{-1}
\!\! \bm{S}_{\pil}^{\herm} \, 
\widetilde{\bm{B}}_{\pil}^{\herm}(\thetain) \, 
\bm{r} \: .
\label{eq:beta-ls-general}
\ee
The LS solution is unique provided that the 
following condition is satisfied: 
{\bf (c1)} the matrix $\widetilde{\bm{B}}_{\pil}(\thetain)$
has full column rank for every $\thetain\in\Theta_\text s$, i.e.,
$\rank[\widetilde{\bm{B}}_{\pil}(\thetain)] =P$,
$\forall\,\thetain\in\Theta_\text s$.
Substituting \eqref{eq:beta-ls-general} into
\eqref{eq:joint-ml-theta-beta} yields 
the concentrated ML estimator
\be
\widehat{\theta}_\text{s}
=
\arg\min_{\thetain\in\Theta_\text s}
\left\|
\bm{P}_{\pil}^{\perp}(\thetain) \, \bm{r}
\right\|^2,
\label{eq:theta_concentrated_ml_min}
\ee
where
$
\bm{P}_{\pil}(\thetain)
\eqdef
\widetilde{\bm{B}}_{\pil}(\thetain)
[
\widetilde{\bm{B}}_{\pil}^\herm(\thetain)
\widetilde{\bm{B}}_{\pil}(\thetain)
]^{-1}
\widetilde{\bm{B}}_{\pil}^\herm(\thetain)
$
is the orthogonal projector onto the column space of
$\widetilde{\bm{B}}_{\pil}(\thetain)$, and
$
\bm{P}_{\pil}^{\perp}(\thetain)
\eqdef
\bm{I}_{M} - \bm{P}_{\pil}(\thetain)
$
is the projector onto its orthogonal complement.
Equivalently, the concentrated ML estimator 
can be expressed as
\be
\widehat{\theta}_\text{s}
=
\arg\max_{\thetain\in\Theta_\text s} 
\bm{r}^{\herm} \, \bm{P}_{\pil}(\thetain) \, \bm{r} \: .
\label{eq:theta-concentrated-ml-max}
\ee 
Once $\widehat{\theta}_{\text{s}}$ has been obtained, the
corresponding composite channel estimate is calculated
as $\widehat{\boldsymbol{\beta}}_{\pil} = 
\widehat{\boldsymbol{\beta}}_{\pil} ( \widehat{\theta}_{\text{s}} )$.

Note that~\eqref{eq:theta-concentrated-ml-max} is nondegenerate since $P<M$
by~\eqref{eq:M-bounds}: if $P=M$, condition~\textbf{(c1)} would make
$\widetilde{\mathbf B}_{\pil}(\thetain)$ square and nonsingular,
so that $\mathbf P_{\pil}(\thetain)=\mathbf I_M$ and the
objective would be independent of $\thetain$. 
Hence, the VCs do not serve as spectral guards, as in conventional OFDM systems, but leave room
for the STCM-generated harmonics, which provide the additional
observations needed to estimate the AoA together with the channel.

While condition \textbf{(c1)} ensures a unique channel estimate
for each candidate AoA, AoA identifiability further requires
different AoAs to generate distinct signal subspaces. 
This leads to the following condition:
\textbf{(c2)} For every distinct pair
$\theta_1,\theta_2\in\Theta_{\text{s}}$,
$
\text{range}
[\widetilde{\bm{B}}_{\pil}(\theta_1)]
\neq
\text{range}
[\widetilde{\bm{B}}_{\pil}(\theta_2)].$
Condition \textbf{(c2)} prevents two distinct AoAs from producing the
same concentrated ML objective for every possible observation vector.
This condition depends jointly on the selected pilot subcarriers and the STCM
coding patterns.

During the acquisition phase, the STCM coding matrix is generated
\textit{a priori} and is known to both the STCM controller and the RX,
together with a calibrated model of the meta-atom responses associated
with the available control states.
Its entries can be drawn among the available control states by a
pseudo-random generator with a common seed, and the resulting realization
is assumed to satisfy \textbf{(c1)} and \textbf{(c2)}.

\subsection{Channel reconstruction over the full OFDM bandwidth}
\label{subsec:parametric-reconstruction}

The estimate $\widehat{\boldsymbol{\beta}}_{\text{pil}}(\widehat{\theta}_{\text{s}})$ 
provides the composite channel coefficients only on the
$P$ active pilot subcarriers.
Under the adopted parametric propagation model, they are
defined over all OFDM subcarriers as
\be
\beta_k
=
\ku
\sum_{\ell=1}^{\Ls}
g_{\rm r} \, g_{{\rm s},\ell} \, 
e^{-j2\pi f_k\tau_{{\rm s},\ell}},
\qquad
\forall k\in\mathcal{O}_{\Nc}.
\label{eq:parametric_composite_channel}
\ee
Reconstructing the channel frequency response over the whole OFDM
bandwidth from pilot-domain samples is a well-established problem,
commonly addressed through either parametric estimation
or interpolation-based methods~\cite{StoicaMoses2005,Coleri2002}.
Among the possible alternatives, we adopt here the parametric
\textit{matrix-pencil} method~\cite{HuaSarkar1990}, owing to its favorable
trade-off between computational complexity and estimation accuracy; it
exploits the multipath structure in
\eqref{eq:parametric_composite_channel} to estimate the path delays and
the associated effective coefficients.
The corresponding channel samples can then be expressed as
\be
\beta_{k_p}
=
\sum_{\ell=1}^{\Ls}
c_\ell \,  z_\ell^p \:,
\label{eq:pilot-exponential-model}
\ee
where $z_\ell \eqdef e^{-j\frac{2\pi}{\Tu}D\tau_{\text{s},\ell}}$ and
$c_\ell \eqdef \ku\,g_{\text r}\,g_{\text{s},\ell}\,
e^{-j2\pi f_{k_0}\tau_{\text{s},\ell}}$.
Collecting the $P$ pilot-domain channel samples yields the compact Vandermonde matrix
representation
\be
\boldsymbol{\beta}_{\pil}
=
\text{Vand}_P(\bm{z})\,\bm{c},
\label{eq:pilot-vandermonde-model}
\ee
with $\bm{c}
\eqdef
[c_1,c_2,\ldots,c_{\Ls}]^\trasp \in\Cset^{\Ls}$ and $\bm{z}
\eqdef
[z_1,z_2,\ldots,z_{\Ls}]^\trasp \in\Cset^{\Ls}$.
The matrix-pencil method recovers the $\Ls$ complex exponentials
$\{z_\ell\}_{\ell=1}^{\Ls}$ from
$\widehat{\boldsymbol{\beta}}_{\text{pil}}$ by constructing two shifted
Hankel matrices from its entries and extracting the $\Ls$-dimensional
signal subspace through a truncated SVD (the detailed derivation is reported, for completeness, in the supplementary material).
Its applicability requires a pencil parameter $J$ such that
\be
\Ls \leq J \leq P-\Ls \:,
\label{eq:pencil-cond}
\ee
where the channel order $\Ls$ is assumed known at the
receiver.\footnote{When unknown, $\Ls$ can be determined using standard
channel-order estimation techniques.}
From the resulting estimates $\widehat{\bm{z}} \eqdef
[\widehat z_1,\ldots,\widehat z_{\Ls}]^\trasp$, the delay estimates follow
as
\be
\widehat{\tau}_{\text{s},\ell}
=
\frac{\Tu}{2\pi D} \, 
\text{mod}
\left(
-\arg\widehat z_\ell,2\pi
\right),
\label{eq:matrix-pencil-delays}
\ee
for $\ell \in \mathcal{J}_{\Ls}$, whereas the effective coefficients are obtained by
least squares from \eqref{eq:pilot-vandermonde-model} as
$\widehat{\bm{c}} = \left[\text{Vand}_P(\widehat{\bm{z}})\right]^\dagger
\widehat{\boldsymbol{\beta}}_{\text{pil}}$.
The frequency response of the composite channel 
is finally reconstructed on all
OFDM subcarriers, for $k \in \mathcal{O}_{\Nc}$, as
\be
\betahat_k
=
\sum_{\ell=1}^{\Ls}
\widehat c_\ell \, 
e^{-j\frac{2\pi}{\Tu}(k-k_0)
\widehat{\tau}_{\text{s},\ell}} .
\label{eq:matrix-pencil-reconstruction}
\ee
The matrix-pencil construction provides a direct criterion for
selecting the number $P$ of active pilot subcarriers.
First, the pencil parameter $J$ must satisfy
\eqref{eq:pencil-cond},
thus a feasible value of $J$ exists only if
$P \geq 2 \, \Ls$.
Therefore, in the noiseless case, $P=2\Ls$ uniformly spaced
channel samples are sufficient to identify $\Ls$ distinct
complex exponential components. In the presence of noise,
choosing $P>2\Ls$ provides additional observations and generally
improves the estimation of the underlying $\Ls$-dimensional
signal subspace~\cite{HuaSarkar1990}.
A second constraint arises from delay ambiguity. Indeed, the
mapping from $\tau_{{\text{s}},\ell}$ to $z_\ell$ in
\eqref{eq:pilot-exponential-model} is periodic with period
$\Tu/D$. 
Thus, unambiguous estimation over
$0 \leq \tau_{\text{s},\ell} \leq \Delta\tau_{\text{s},\text{max}}$ requires
$\Delta\tau_{\text{s},\text{max}} < \Tu/D$.
Since $D\leq(\Nc-1-2H)/(P-1)$ by~\eqref{eq:pilot-spacing} and
$\Tu=\Nc\Tc$, a sufficient condition is
$\Delta\tau_{\text{s},\text{max}} < \Nc\Tc\,(P-1)/(\Nc-1-2H)$.
Combining the matrix-pencil identifiability and delay-unambiguity
requirements yields the following lower bound:
\be
P \geq P_{\text{min}}
\eqdef
\max
\left\{ 2\,\Ls,\,
\left\lfloor
\frac{\Delta\tau_{\text{s},\text{max}}}{\Tc}\,
\frac{(\Nc-1-2H)}{\Nc}
\right\rfloor+2
\right\}.
\label{eq:number-pilots}
\ee


\subsection{Time-invariant RIS configuration and OFDM demodulation}
\label{sec:OFDM-detection}

During the communication phase, all $\Nc$ subcarriers
carry data symbols, with no acquisition-specific pilots
or VCs being introduced.%
\footnote{Conventional pilots or VCs employed
by the underlying OFDM system are not explicitly included in the
model.}
Moreover, the temporal modulation at the STCM is disabled, so that the surface
operates as a conventional time-invariant RIS.
Consequently, no modulation-induced harmonics 
are generated, which eliminates ICI and
considerably simplifies OFDM reception (see
Remark~\ref{rem:structured-ICI}).
Each STCM column is assumed to be 
driven by a constant control signal, so that the $q$-th
meta-atom remains in a fixed control state 
$\chi_q$, for $q \in \mathcal{J}_{\Qc}$.
The state vector
$\boldsymbol{\chi} \eqdef [\chi_1,\chi_2,\ldots,\chi_{\Qc}]^\trasp$
therefore represents the design variable for the communication phase.
The set of admissible control
states is denoted by $\mathcal{X}$.
This set is continuous when the
meta-atoms are continuously tunable, whereas it contains $2^b$
states under $b$-bit quantized control.

The physics-based model developed in
Section~\ref{sec:meta-atom-model} then reduces to its
time-invariant counterpart,
$\Gamma_q(f,t;\thetain)=\Gamma_q(f;\thetain,\chi_q)$.
This response is obtained from the pole--residue representation
\eqref{eq:pole-residue} by fixing the poles, residues, and direct
term to the values associated with the applied control state
$\chi_q$.
The dependence on $\chi_q$ is made explicit because the control
states constitute the design variables for the data-transmission
phase.

By specializing \eqref{eq:yh_ofdm_scalar} to the time-invariant case, only
the term with $k=h$ remains. Hence
\be
r_{h}[n]  = H_{h}(\boldsymbol{\chi}) \, s_{h}[n] + v_{h}[n],
\label{eq:yh-data}
\ee
where, recalling the definition of $\beta_h$,
\be
H_{h}(\boldsymbol{\chi}) \eqdef \beta_{h}
\sum_{q=1}^{\Qc} \Gamma_q(f_h;\thetain, \chi_q )
\, e^{-j2\pi(q-1)\left(1+\frac{f_h}{f_c}\right)\left(\psi_{\text{s}}+\psi_{\text{r}}\right)} 
\label{eq:effective-channel}
\ee
denotes the end-to-end channel 
experienced by the $h$-th subcarrier, for
$h\in \mathcal{O}_{\Nc}$ 
and an arbitrary
control-state vector $\boldsymbol{\chi}$.
Equation~\eqref{eq:yh-data} confirms that, in the communication phase,
the OFDM subcarriers can be processed independently by the RX.

The AoA estimate $\widehat{\theta}_{\text{s}}$ 
obtained during the acquisition phase is exploited to configure the STCM for the communication
phase.
Since the AoD toward the RX is assumed to be known, the corresponding
normalized spatial frequency $\psi_{\text{r}}$ is available.
Since the STCM configuration $\boldsymbol{\chi}$ is common to all
subcarriers, whereas the configuration maximizing $|H_h(\boldsymbol{\chi})|$
generally differs from one subcarrier to another, a design
criterion that jointly accounts for all subcarriers is needed.
We propose here to select the static control states 
according to the following
\textit{max--min criterion:}
\begin{multline}
\widehat{\boldsymbol{\chi}}
=
\arg\max_{\boldsymbol{\chi} \in \mathcal{X}^{\Qc}}
\min_{h \in \mathcal{O}_{\Nc}}
\left|
\sum_{q=1}^{\Qc}
\Gamma_q (f_h;\widehat{\theta}_{\text{s}}, \chi_q ) \right. \\
\times
\left. 
e^{-j 2\pi (q-1)\left(1+\frac{f_h}{f_c}\right)
(\widehat{\psi}_{\text{s}} + \psi_{\text{r}})}
\right|,
\label{eq:minimax-design}
\end{multline}
where $\widehat{\psi}_{\text{s}} \eqdef
\lambda_c^{-1} d \, \sin \widehat{\theta}_{\text{s}}$.
The adopted criterion maximizes the coherent combining gain of the weakest
OFDM subcarrier, thereby avoiding severe gain degradation on any
subcarrier.

Problem~\eqref{eq:minimax-design} is generally nonconvex; nevertheless,
each control state $\chi_q$ affects the objective function 
only through the corresponding reflection response
$\Gamma_q(f_h;\widehat{\theta}_{\text{s}},\chi_q)$.
This structure enables a \textit{block-coordinate ascent} (BCA) implementation, in
which the states are cyclically updated by solving one-dimensional
subproblems while keeping the remaining ones fixed~\cite{Bertsekas2016}.
For a $b$-bit quantized meta-atom control, 
each subproblem is solved exactly by
enumerating the $2^b$ admissible states.
Although this procedure does not generally guarantee convergence to a
globally optimal configuration, it produces a nondecreasing sequence of
objective values and offers a computationally tractable solution.

Finally, data detection is performed through conventional one-tap
zero-forcing frequency-domain OFDM 
equalization $\widehat{s}_{h}[n] = \widehat{H}_{h}^{-1} \, r_{h}[n] $,
$h \in \mathcal{O}_{\Nc}$, followed by minimum-distance detection over the
signal constellation.
Here, the channel estimate $\widehat{H}_{h}$ is obtained from
\eqref{eq:effective-channel} by replacing $\beta_{h}$ and $\thetain$ with
their estimates 
obtained in the acquisition phase 
and using the optimized control states
$\widehat{\boldsymbol{\chi}}$.

\begin{remark}[Signaling requirements]
\label{rem:signaling}
The proposed protocol requires the TX, the RX, and the STCM controller to
be synchronized, as discussed in Section~\ref{sec:dual-mode-protocol},
whereas sharing the coding matrix used in the acquisition phase only
requires a common pseudo-random seed (see
Section~\ref{subsec:vc-aided-acquisition}).
In the communication phase, the RX can feed back only the quantized AoA
estimate, in which case the STCM controller solves~\eqref{eq:minimax-design}
with a BCA complexity of $\mathcal{O}(\Qc\,2^b\Nc)$ operations per sweep.
Alternatively, optimization in \eqref{eq:minimax-design}
can be performed at the RX, which then
feeds back the $\Qc b$ bits of $\widehat{\boldsymbol{\chi}}$, possibly
through limited-feedback schemes~\cite{Love2008}.
\end{remark}

\section{Numerical results}
\label{sec:simulations}

In this section, we assess the proposed framework through numerical
simulations.
In particular, Section~\ref{sec:metaatom-results} characterizes the
meta-atom model proposed in Section~\ref{sec:meta-atom-model}, showing
its significant frequency selectivity and spatial dispersion over the
signal band.
In Section~\ref{sec:isac-results}, we evaluate the performance of the
dual-mode protocol introduced in 
Section~\ref{sec:dual-mode-protocol}, in terms of AoA
estimation accuracy and spectral efficiency, also in comparison with the
wideband channel estimator of~\cite{Ma2021}.

We consider an OFDM system operating in the 5G NR band n78, with carrier
frequency $f_c=3.5$~GHz, SCS $\Delta f=30$~kHz
($\Tu=33.33~\mu$s), and $\Nc=3276$ subcarriers (273 resource blocks),
which occupy a bandwidth $\Bs=\Nc \Delta f=98.28$~MHz centered on $f_c$.
Choosing a normal CP, with $\Tcp=2.34~\mu$s, 
yields $\ku=0.934$ and a  
sampling period $\Tc=\Tu/\Nc=10.2$~ns.
The STCM consists of $\Qc=32$ columns of the varactor-loaded unit cells
designed in~\cite{Rains2023}, 
with period $d=0.35\,\lambda_c = 30$~mm.
In the acquisition phase, the modulation period $T=\Tu$ is divided into
$K=64$ slots of duration $\Tr=\Tu/K=520.8$~ns.

Each meta-atom is described by an equivalent-circuit admittance model,
which yields the reflection coefficient in closed form for each AoA value
and polarization (TE or TM); for completeness, the model is detailed in
the supplementary material.
Four circuit parameters are not available in closed form, namely, the
bias shunt capacitance, the via inductance, the patch-to-pad inductance,
and the varactor series resistance at $f_c$.
They are calibrated on the measured 1-bit response reported
in~\cite{Rains2023} for the control voltages $2.5$~V and $10$~V, 
i.e., for a phase difference between the two states of $200^\circ$ at $3.5$~GHz and
of $160^\circ$ at $3.42$ and $3.58$~GHz, and a maximum reflection loss of
$4.2$~dB at $3.42$~GHz.%
\footnote{As an independent check, between $1$~V
and $10$~V the calibrated model gives a phase range of $242^\circ$ at
$3.5$~GHz, against $248^\circ$ in the full-wave simulation
of~\cite{Rains2023}.}
The remaining circuit parameters of the unit cell
follow from the cell geometry itself, the substrate, and the
varactor datasheet.
The meta-atoms are 1-bit devices, i.e.,
$\mathcal{X}=\{\chi^{(0)},\chi^{(1)}\}$, whose control states are
obtained with reverse bias voltages of $1.65$~V and $7.19$~V,
respectively; the choice of the operating band and the two states is
further discussed in Section~\ref{sec:metaatom-results}.
Unless otherwise stated, the incident field is TE-polarized, and the
admissible AoA range is  $\Theta_\text{s}=[-85^\circ,85^\circ]$.

\subsection{Meta-atom model: spatial dispersion and frequency selectivity}
\label{sec:metaatom-results}

The operating frequency and the two control states 
are chosen within the $3.3$--$3.7$~GHz
portion of the n78 band
so that the meta-atom behaves as an effective 1-bit
element over the signal band, i.e., with the two states close to phase
opposition and with low losses.
The two states are optimized at normal incidence ($\thetain = 0$) 
in two steps: for each candidate center frequency, 
the bias voltages maximize the worst-case
steerable gain $|\Gamma_q(f;0,\chi^{(1)})-\Gamma_q(f;0,\chi^{(0)})|$ over
the signal band, which accounts for both phase difference and losses; then,
$f_c=3.5$~GHz is the candidate that minimizes the worst-case deviation of
$\arg[\Gamma_q(f;0,\chi^{(1)})/\Gamma_q(f;0,\chi^{(0)})]$ from $180^\circ$.
For $f\in(-\Bs/2,\Bs/2)$, the resulting phase difference ranges between
$154^\circ$ and $213^\circ$ ($197^\circ$ at $f_c$), and the magnitude of
both states stays above $-2.3$~dB.

\begin{table}[t]
\centering
\caption{Direct term $\Gamma_\infty$, 
poles $\rho_p$, and zeros $\xi_k$ of the meta-atom model
in \eqref{eq:pole-residue}
at normal incidence ($\rho_p/2\pi$ and $\xi_k/2\pi$ in GHz).}
\label{tab:poles}
\footnotesize
\setlength{\tabcolsep}{4pt}
\begin{tabular}{@{}llcc@{}}
\toprule
 & & $\chi^{(0)}$ ($1.65$~V) & $\chi^{(1)}$ ($7.19$~V) \\
\midrule
\multicolumn{2}{@{}l}{$\Gamma_{\infty}$} & $1.00$ & $1.00$ \\
\midrule
 & real & $-212.86$ & $-212.82$ \\
$\rho_p$ & far pair & $-0.034\pm j4.423$ & $-0.068\pm j4.844$ \\
 & dominant pair & $-0.113\pm j3.326$ & $-0.102\pm j3.563$ \\
\midrule
 & real & $212.46$ & $212.41$ \\
$\xi_k$ & far pair & $-0.033\pm j4.424$ & $-0.027\pm j4.850$ \\
 & dominant pair & $0.061\pm j3.329$ & $0.077\pm j3.563$ \\
\bottomrule
\end{tabular}
\end{table}

The resulting rational model in 
\eqref{eq:pole-residue} has $\Np=5$ poles, i.e., one real pole and
two complex-conjugate pairs, which are reported in Table~\ref{tab:poles}
at normal incidence together 
with the $\Nz = 5$ 
zeros $\xi_k$ and the direct term $\Gamma_\infty$.
The residues in~\eqref{eq:pole-residue} follow from these values as
$\zeta_p=\Gamma_{\infty}\prod_{k=1}^{\Nz}(\rho_p-\xi_k)/
\prod_{k\neq p}(\rho_p-\rho_k)$.
Dropping the indices $q$ and $p$ hereafter, we discuss the three groups of poles separately.
The real pole lies at $\rho/2\pi=-212.86$~GHz, and its residue is
$\zeta\approx2\rho$, since the real zero is nearly opposite to it
($212.46$~GHz in Table~\ref{tab:poles}).
By~\eqref{eq:lowpass-meta-atom-impulse-response}, its contribution to the
impulse response decays with time constant $1/|\rho|\approx0.75$~ps,
far shorter than $\Tc = 10.2$~ns, and thus acts as an impulse of area
$\zeta/(j2\pi f_c-\rho)\approx-\zeta/\rho\approx-2$.
Added to $\Gamma_{\infty}=1$, it yields an impulse area
$\approx-1$, i.e., the reflection coefficient of the ground
plane, on top of which the resonant terms act.
The far pair, at $4.42$~GHz ($\chi^{(0)}$) and $4.84$~GHz ($\chi^{(1)}$),
about $1$~GHz above the signal band, is nearly 
cancelled by the far zeros, lying at almost the same frequencies:
evaluating~\eqref{eq:pole-residue-f} over the signal band, its 
contribution does not exceed $0.03$ in magnitude, against at least $0.72$ for
the dominant pair.
The latter, at $3.33$~GHz ($\chi^{(0)}$) and $3.56$~GHz
($\chi^{(1)}$), is the resonance that makes the surface reconfigurable;
the signal band lies between the two resonance frequencies, where the two
phase responses are approximately in opposition.

\begin{figure}[t]
\centering
\includegraphics[width=0.95\columnwidth]{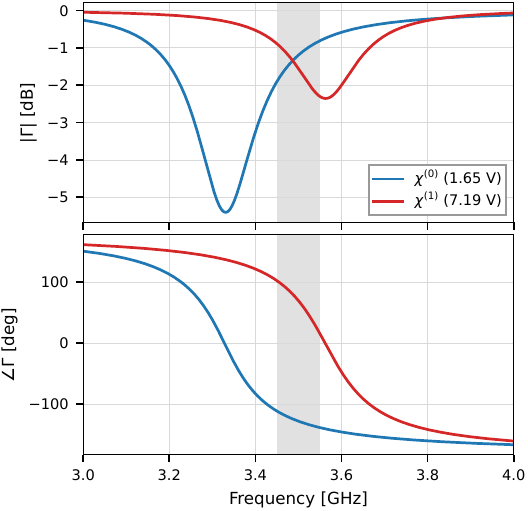}
\caption{Magnitude (top) and phase (bottom) of the reflection coefficient
of the meta-atom in the two control states versus frequency, at normal incidence. The
shaded area is the considered signal band.}
\label{fig:meta-freq}
\end{figure}

Over $\Theta_\text{s}$ and both polarizations, the slowest-decaying mode is
the far pair of $\chi^{(0)}$ in TE at the edge of $\Theta_\text{s}$,
which, accounting for~\eqref{eq:Tgamma} with $\epsilon=10^{-2}$,
yields $\Tgamma=28.7$~ns.
This value is conservative, since~\eqref{eq:rho-max} does not weigh the
modes by their residues, and the far pair, being almost cancelled by
nearby zeros, has small residues.
Since $\Tgamma$ is about $18$ times shorter than $\Tr = 520.8$~ns, 
the adiabatic approximation holds; moreover, since $\Tgamma+2(\Qc-1)d/c=34.9$~ns, the
CP condition~\eqref{eq:CP-cond} leaves more than $2.3~\mu$s 
for accommodating the delay spread of the TX--STCM channel.
However, $\Bs \Tgamma\approx 2.8$, i.e., the meta-atom operates in the
frequency-selective regime of the first row of
Table~\ref{tab:Gamma-models}.
Indeed, results of Fig.~\ref{fig:meta-freq} show that 
the phases of $\chi^{(1)}$ and
$\chi^{(0)}$ vary within the band by $85^\circ$ and $27^\circ$, and their
magnitudes by $1.4$~dB and $1.1$~dB, respectively, 
so that the frequency-flat model of
the second row would misrepresent the response away from broadside, i.e., over most of the AoA range to be estimated.

\begin{figure}[t]
\centering
\includegraphics[width=0.95\columnwidth]{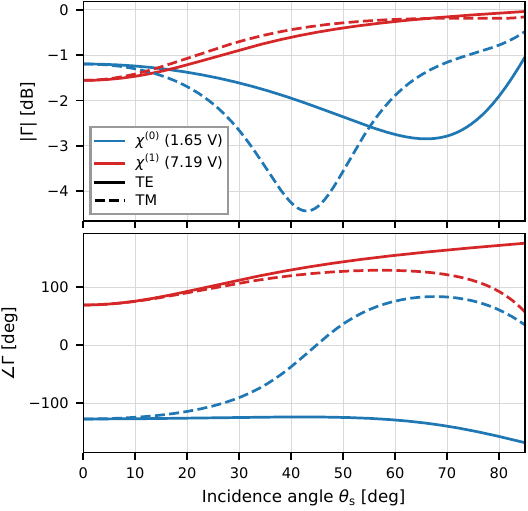}
\caption{Magnitude (top) and phase (bottom) of the reflection coefficient
of the meta-atom in the two control states at $f_c = 3.5$~GHz versus the AoA, for TE (solid) and TM
(dashed) polarization.}
\label{fig:meta-angle}
\end{figure}

Fig.~\ref{fig:meta-angle} shows the spatial dispersion of the reflection coefficient  
at $f_c$.
The angular dependence is significant already at moderate angles and
differs between TE and TM: at $60^\circ$, the phase difference between
the two states 
moves from $197^\circ$ to $285^\circ$ in TE and  to $53^\circ$ in TM, far from phase opposition in both cases.
Hence, the pair optimized at normal incidence does not keep the two
states in phase opposition for every AoA.
Different pairs could be selected in the communication phase, once the AoA has been estimated; for simplicity, the same pair is used in both phases.
%
%
Finally, adopting the spatially non-dispersive model
in the last row of Table~\ref{tab:Gamma-models},
with $\bar{\theta}_{\text{s}}=0$,
would neglect the dependence of the STCM response
on the AoA, which is precisely the parameter
to be estimated during acquisition.

\subsection{Acquisition and communication performance}
\label{sec:isac-results}

In the acquisition phase, we use $P=40$ pilots with spacing
$D=\bar{D}=82$ and $2H+1=63$ harmonics per pilot.
The TX--STCM channel has $\Ls=4$ paths
with exponentially decaying power, 
normalized to unit total
power, whose delays are randomly 
drawn in $[0,300]$~ns according to a uniform distribution; 
the SNR is $1/\sigma_v^2$, which represents 
the per-subcarrier SNR before the
STCM combining gain, with $|s_{k_p}|=1$.
Results in terms of root-mean-square error (RMSE)
are averaged over $1000$ Monte Carlo trials.
As a benchmark, we consider the TS-OMP estimator of~\cite{Ma2021}, which
accounts for beam squint but models the RIS elements as ideal
unit-modulus, frequency- and angle-independent phase shifters.
Following~\cite{Ma2021}, angular diversity is obtained by reconfiguring
the RIS
over $16$ pilot OFDM symbols, with the same total pilot energy as
the proposed scheme.
We also report the performance  
of the proposed estimator when beam squint is
neglected,
i.e., under the narrowband-array approximation
discussed in
Section~\ref{subsec:ofdm-harmonic}, together with the Cram\'er--Rao lower
bound (CRLB) on the AoA.

\begin{figure}[t]
\centering
\includegraphics[width=0.95\columnwidth]{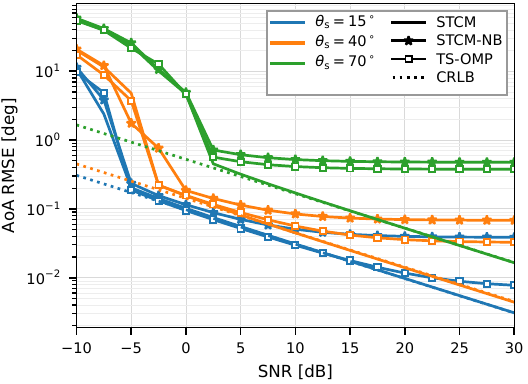}
\caption{AoA RMSE versus SNR for 
$\thetain=15^\circ$, $40^\circ$, and
$70^\circ$: 
proposed estimator (STCM, $1$ pilot OFDM symbol), 
proposed estimator with beam squint neglected 
(STCM-NB, $1$ pilot OFDM symbol),
TS-OMP of~\cite{Ma2021} ($16$ pilot OFDM symbols), 
and CRLB.}
\label{fig:aoa}
\end{figure}

Fig.~\ref{fig:aoa} shows the AoA  RMSE versus the
SNR, for $\thetain=15^\circ$, $40^\circ$, and $70^\circ$.
All estimators exhibit the threshold behavior typical of nonlinear
estimation: below an SNR value, which increases with $\thetain$
(about $-5$, $-2.5$, and $2.5$~dB, respectively), outliers dominate and
the RMSE significantly departs from the CRLB.
Above threshold, the proposed estimator attains the CRLB for all the
considered AoAs, even using a single pilot OFDM symbol.
By contrast, the RMSE of TS-OMP, which uses $16$ times more pilot symbols at the same total energy,
saturates at high SNR, with an error floor that grows with the AoA.
This floor is caused by model mismatch: TS-OMP neglects both the
frequency selectivity and the spatial dispersion of the meta-atom
response, the latter becoming increasingly relevant; at high SNR, this mismatch dominates over the noise.
The RMSE of STCM-NB also saturates at high SNR but for a different
reason: although it accounts for the meta-atom dispersion, it neglects
beam squint, which confirms the need to model it for the considered
signal bandwidth and STCM aperture.
Simulations not reported here show that the estimator of the composite
channel on the pilot subcarriers attains the CRLB, whereas the
matrix-pencil reconstruction exhibits a degradation of about $0.5$~dB
with respect to the CRLB, consistently with~\cite{HuaSarkar1990}.

\begin{figure}[t]
\centering
\includegraphics[width=0.95\columnwidth]{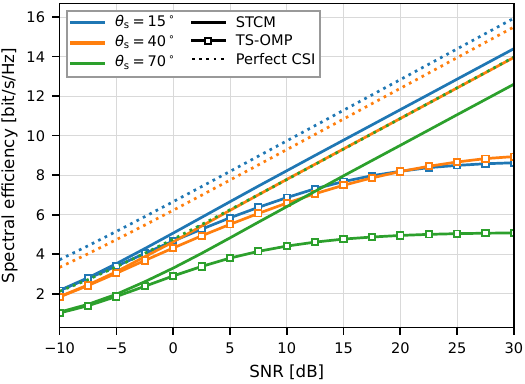}
\caption{Spectral efficiency versus SNR for $\thetain=15^\circ$, $40^\circ$,
and $70^\circ$. The STCM configuration and the data equalization rely on
the AoA and channel estimates provided by the proposed method (STCM) or
by TS-OMP~\cite{Ma2021}, or on the true AoA and channel (perfect CSI).}
\label{fig:rate}
\end{figure}

Fig.~\ref{fig:rate} reports the spectral efficiency $R$, 
averaged over 1000 Monte Carlo trials, defined as
\be
R \eqdef \left(1-\frac{N_\text{acq}}{\Lcoh}\right)\frac{\ku}{\Nc}
\sum_{h\in\mathcal{O}_{\Nc}}\log_2\left(1+
\frac{|\widehat{H}_h|^2}{|H_h-\widehat{H}_h|^2+\sigma_v^2}\right),
\label{eq:rate}
\ee
where $H_h$ and $\widehat{H}_h$ are shorthands for
$H_h(\widehat{\boldsymbol\chi})$ in~\eqref{eq:effective-channel} and for
its estimate obtained by replacing $\beta_h$ and $\thetain$ with their
estimates from the acquisition phase, respectively, with
$\widehat{\boldsymbol\chi}$ given by~\eqref{eq:minimax-design}; the channel estimation
error is treated as additional noise, and $N_\text{acq}$ out of $\Lcoh=140$ OFDM symbols are used for
acquisition, i.e., $N_\text{acq}=1$  ($0.7\%$ overhead) for the proposed
scheme and $N_\text{acq}=16$  ($11.4\%$ overhead) for TS-OMP.
The spectral efficiency of the proposed method grows with the SNR with the
same slope as that achieved with perfect CSI, for all the considered
angles.
By contrast, the spectral efficiency of TS-OMP saturates at high SNR,
at about $9$~bit/s/Hz for $\thetain=15^\circ$ and $40^\circ$ and $5$~bit/s/Hz
for $\thetain=70^\circ$, since the model mismatch yields
channel estimation errors that do not vanish as the SNR increases.

\section{Conclusions}
\label{sec:conclusions}

We have proposed a dual-mode STCM-assisted ISAC framework for OFDM
systems, based on a physics-based model of the meta-atom that accounts for
its frequency selectivity and spatial dispersion, as well as for beam
squint across the STCM aperture.
By aligning the STCM-induced harmonics with the OFDM grid, the AoA and the
composite channel are jointly estimated from a single pilot OFDM symbol,
whereas a static configuration of the surface is used for data
transmission.
Numerical results for a realistic varactor-based unit cell have shown that
the proposed AoA estimator attains the
Cram\'er--Rao bound above the SNR threshold, 
and that the resulting spectral
efficiency grows with the SNR as that achieved with perfect CSI, whereas
it saturates when the meta-atom response is modeled as an ideal phase
shifter.
Future work includes possible extensions to multi-antenna receivers,
multiple users, and multiple AoAs.

\bibliographystyle{IEEEtran}
\bibliography{STC-RIS_bib_v7}

@article{Love2008,
  author  = {Love, David J. and Heath, Robert W. and Lau, Vincent K. N.
             and Gesbert, David and Rao, Bhaskar D. and Andrews, Matthew},
  title   = {An Overview of Limited Feedback in Wireless Communication Systems},
  journal = {IEEE Journal on Selected Areas in Communications},
  volume  = {26},
  number  = {8},
  pages   = {1341--1365},
  month   = oct,
  year    = {2008},
  doi     = {10.1109/JSAC.2008.081002}
}

@article{Luukkonen:2008sa,
  title = {Simple and Accurate Analytical Model of Planar Grids and High-Impedance Surfaces Comprising Metal Strips or Patches},
  author = {Luukkonen, Olli and Simovski, Constantin and Granet, {\relax G{\'e}}rard and Goussetis, George and Lioubtchenko, Dmitri and Raisanen, Antti V. and Tretyakov, Sergei A.},
  year = 2008,
  month = jun,
  journal = {IEEE Transactions on Antennas and Propagation},
  volume = {56},
  number = {6},
  pages = {1624--1632},
  issn = {0018-926X},
  doi = {10.1109/TAP.2008.923327},
  urldate = {2020-05-08},
}

@book{Gardner1990,
  author    = {William A. Gardner},
  title     = {Introduction to Random Processes with Applications to Signals and Systems},
  edition   = {2},
  publisher = {McGraw-Hill},
  address   = {New York},
  year      = {1990},
  isbn      = {9780070225825}
}

@article{Cui2026,
    author = {Dai, Zihao and Zheng, Yi Ning and Zhang, Lei and Chen, Xiao Qing and Castaldi, Giuseppe and Cui, Tie Jun and Galdi, Vincenzo},
    title = {Joint Control of Radiated and Surface Waves via Space-Time Coding Metasurfaces},
    journal = {Advanced Functional Materials},
    volume = {36},
    number = {25},
    pages = {e19200},
    year = {2026}
}

@misc{Mizmizi2026,
  author        = {Mizmizi, Marouan and
                   Tebaldini, Stefano and
                   Spagnolini, Umberto},
  title         = {Echo-Side Integrated Sensing and Communication
                   via Space--Time Reconfigurable Intelligent Surfaces},
  year          = {2026},
  month         = jan,
  eprint        = {2601.09484},
  archiveprefix = {arXiv},
  primaryclass  = {eess.SP},
  doi           = {10.48550/arXiv.2601.09484},
  url           = {https://arxiv.org/abs/2601.09484}
}

@ARTICLE{Ismail2024,
  author={Ismail, Mohamed I. and Shaheen, Abdullah M. and Fouda, Mostafa M. and Alwakeel, Ahmed S.},
  journal={IEEE Open Journal of the Communications Society}, 
  title={{RIS}-Assisted Integrated Sensing and Communication Systems: Joint Reflection and Beamforming Design}, 
  year={2024},
  volume={5},
  number={},
  pages={908-927}}

@ARTICLE{Iudice2026,
      author={Iudice, Ivan and Darsena, Donatella and Gelli, Giacinto and Galdi, Vincenzo},
      journal={	arXiv:2609.18360}, 
      title={Physically consistent modeling of dispersive time-modulated reconfigurable intelligent surfaces for wideband {OFDM}}, 
  year={2026}
}

@book{Kay1993,
  author    = {Steven M. Kay},
  title     = {Fundamentals of Statistical Signal Processing, Volume {I}: Estimation Theory},
  publisher = {Prentice Hall},
  address   = {Upper Saddle River, NJ},
  year      = {1993},
  isbn      = {9780133457118}
}

@Article{Costa2021,
  author  = {F. Costa and M. Borgese},
  journal = {IEEE Open Journal of the Communications Society},
  title   = {Electromagnetic Model of Reflective Intelligent Surfaces},
  year    = {2021},
  pages   = {1577--1589},
  volume  = {2},
}

@article{Bello1963,
  author    = {Philip A. Bello},
  title     = {Characterization of Randomly Time-Variant Linear Channels},
  journal   = {IEEE Transactions on Communication Systems},
  volume    = {11},
  number    = {4},
  pages     = {360--393},
  year      = {1963},
  month     = dec,
  doi       = {10.1109/TCOM.1963.1088793}
}

@ARTICLE{Darsena2024,
  author={Verde, Francesco and Darsena, Donatella and Galdi, Vincenzo},
  journal={IEEE Transactions on Communications}, 
  title={Rapidly Time-Varying Reconfigurable Intelligent Surfaces for Downlink Multiuser Transmissions}, 
  year={2024},
  volume={72},
  number={6},
  pages={3227-3243},
  doi={10.1109/TCOMM.2024.3358954}
}

@ARTICLE{Ma2021,
  author={Ma, Siqi and Shen, Wenqian and An, Jianping and Hanzo, Lajos},
  journal={IEEE Transactions on Wireless Communications}, 
  title={Wideband Channel Estimation for {IRS}-Aided Systems in the Face of Beam Squint}, 
  year={2021},
  volume={20},
  number={10},
  pages={6240-6253},
  doi={10.1109/TWC.2021.3072694}
}

@book{Oppenheim1997Signals,
  author    = {Oppenheim, Alan V. and Willsky, Alan S. and Nawab, S. Hamid},
  title     = {Signals and Systems},
  edition   = {2nd},
  publisher = {Prentice Hall},
  address   = {Upper Saddle River, NJ, USA},
  year      = {1997}
}

@article{Minkov2017Exact,
  author  = {Minkov, Momchil and Shi, Yu and Fan, Shanhui},
  title   = {Exact Solution to the Steady-State Dynamics of a Periodically Modulated Resonator},
  journal = {APL Photonics},
  volume  = {2},
  number  = {7},
  pages   = {076101},
  month   = jul,
  year    = {2017},
  doi     = {10.1063/1.4985381}
}

@article{Ghasemi2021,
  author  = {Ghasemi, Sepideh and Rajabalipanah, Hamid and Tayarani, Majid and Abdolali, Ali and Baharian, Mohammad},
  title   = {Time-Modulated Metasurface-Assisted Measurements},
  journal = {Advanced Optical Materials},
  volume  = {9},
  number  = {7},
  pages   = {2001594},
  year    = {2021},
  doi     = {10.1002/adom.202001594}
}

@article{Bjornson2017,
  author  = {Emil Bj{\"o}rnson and Jakob Hoydis and Luca Sanguinetti},
  title   = {Massive {MIMO} Networks: Spectral, Energy, and Hardware Efficiency},
  journal = {Foundations and Trends in Signal Processing},
  year    = {2017},
  volume  = {11},
  number  = {3--4},
  pages   = {154--655},
  doi     = {10.1561/2000000093}
}

@Article{Schmidt1986,
  author   = {Schmidt, R.},
  journal  = {IEEE Transactions on Antennas and Propagation},
  title    = {{Multiple emitter location and signal parameter estimation}},
  year     = {1986},
  number   = {3},
  pages    = {276--280},
  volume   = {34},
  doi      = {10.1109/TAP.1986.1143830},
}

@Article{Roy1989,
  author   = {Roy, R. and Kailath, T.},
  journal  = {IEEE Transactions on Acoustics, Speech, and Signal Processing},
  title    = {{ESPRIT-estimation of signal parameters via rotational invariance techniques}},
  year     = {1989},
  number   = {7},
  pages    = {984--995},
  volume   = {37},
  doi      = {10.1109/29.32276},
}

@Article{Krim1996,
  author  = {H. Krim and M. Viberg},
  journal = {IEEE Signal Processing Magazine},
  title   = {Two Decades of Array Signal Processing Research},
  year    = {1996},
  number  = {4},
  pages   = {67--94},
  volume  = {13},
}

@Article{Cui2014,
  author  = {Cui, Tie Jun and Qi, Meng and Wan, Xiang and Zhao, Jie and Cheng, Qiang},
  journal = {Light: Science \& Applications},
  title   = {Coding Metamaterials, Digital Metamaterials and Programmable Metamaterials},
  year    = {2014},
  pages   = {e218},
  volume  = {3},
  doi     = {10.1038/lsa.2014.99},
}

@Article{Chen2024,
  author   = {Chen, Zhimin and Chen, Peng and Guo, Ziyu and Zhang, Yudong and Wang, Xianbin},
  journal  = {IEEE Transactions on Intelligent Transportation Systems},
  title    = {A {RIS}-Based Vehicle {DOA} Estimation Method With Integrated Sensing and Communication System},
  year     = {2024},
  number   = {6},
  pages    = {5554--5566},
  volume   = {25},
  doi      = {10.1109/TITS.2023.3330172},
}

@article{Chen2025,
  author  = {Chen, Xiao Qing and
             Zhang, Lei and
             Zheng, Yi Ning and
             Liu, Shuo and
             Huang, Zhuo Ran and
             Liang, Jing Cheng and
             {Di Renzo}, Marco and
             Galdi, Vincenzo and
             Cui, Tie Jun},
  title   = {Integrated Sensing and Communication Based on
             Space-Time-Coding Metasurfaces},
  journal = {Nature Communications},
  year    = {2025},
  volume  = {16},
  number  = {1},
  pages   = {1836},
  month   = feb,
  doi     = {10.1038/s41467-025-57137-6}
}

@Article{He2020,
  author  = {He, Zhiqiang and Zhang, Xiaoming and others},
  journal = {IEEE Wireless Communications Letters},
  title   = {Cascaded Channel Estimation for Large Intelligent Metasurface Assisted Massive {MIMO}},
  year    = {2020},
  number  = {2},
  pages   = {210--214},
  volume  = {9},
  doi     = {10.1109/LWC.2019.2948631},
}

@Article{Yu2025,
  author   = {Yu, Canping and Li, Yingsong and Li, Liping and Huang, Zhixiang and Wu, Qingqing and de Lamare, Rodrigo},
  journal  = {IEEE Transactions on Aerospace and Electronic Systems},
  title    = {Dual {Lawson} Norm-Based Robust {DOA} Estimation for {RIS}-Aided Wireless Communication Systems},
  year     = {2025},
  number   = {1},
  pages    = {582--592},
  volume   = {61},
  doi      = {10.1109/TAES.2024.3446752},
}

@Article{Chen2021,
  author   = {Chen, Peng and Yang, Zihan and Chen, Zhimin and Guo, Ziyu},
  journal  = {IEEE Signal Processing Letters},
  title    = {Reconfigurable Intelligent Surface Aided Sparse {DOA} Estimation Method With Non-{ULA}},
  year     = {2021},
  pages    = {2023--2027},
  volume   = {28},
  doi      = {10.1109/LSP.2021.3116789},
}

@Book{Bertsekas2016,
  author    = {Dimitri P. Bertsekas},
  publisher = {Athena Scientific},
  title     = {Nonlinear Programming},
  year      = {2016},
  edition   = {3rd},
  isbn      = {978-1-886529-05-2},
}

@Book{Hu2024,
  author    = {Hu, X. and Liu, C. and Peng, M. and Zhong, C.},
  publisher = {Springer Nature},
  title     = {Reconfigurable Intelligent Surface-Enabled Integrated Sensing and Communication in 6G},
  year      = {2024},
  address   = {Singapore},
}

@Article{Liu2023,
  author  = {Liu, R. and Li, M. and Luo, H. and Liu, Q. and Swindlehurst, A. L.},
  journal = {IEEE Wireless Communications},
  title   = {Integrated sensing and communication with reconfigurable intelligent surfaces: Opportunities, applications, and future directions},
  year    = {2023},
  number  = {1},
  pages   = {50--57},
  volume  = {30},
}

@ARTICLE{Zhang2026,
  author={Zhang, Di and Cui, Yuanhao and Cao, Xiaowen and Su, Nanchi and Gong, Yi and Liu, Fan and Yuan, Weijie and Jing, Xiaojun and Andrew Zhang, J. and Xu, Jie and Masouros, Christos and Niyato, Dusit and Di Renzo, Marco},
  journal={IEEE Communications Surveys \& Tutorials}, 
  title={Integrated Sensing and Communications Over the Years: An Evolution Perspective}, 
  year={2026},
  volume={28},
  number={},
  pages={5014-5048},
  doi={10.1109/COMST.2026.3655674}
}

@article{Zhang2018,
  author  = {Lei Zhang and Xiao Qing Chen and Shuo Liu and Qian Zhang and Jie Zhao and Jun Yan Dai and Guo Dong Bai and Xiang Wan and Qiang Cheng and Giuseppe Castaldi and Vincenzo Galdi and Tie Jun Cui},
  title   = {Space-time-coding digital metasurfaces},
  journal = {Nature Communications},
  volume  = {9},
  number  = {1},
  pages   = {4334},
  year    = {2018},
  doi     = {10.1038/s41467-018-06802-0}
}

@Article{Santos2024,
  author  = {Herman L. dos Santos and Martin Voigt Vejling and Taufik Abr{\~a}o and Petar Popovski},
  journal = {IEEE Transactions on Wireless Communications},
  title   = {Assessing the Potential of Space-Time-Coding Metasurfaces for Sensing and Localization},
  year    = {2024},
  number  = {12},
  pages   = {19397--19411},
  volume  = {23},
  doi     = {10.1109/TWC.2024.3483149},
}

@article{Li2026,
  author  = {Li, Shunyu and
             Mao, Tianqi and
             Liu, Guangyao and
             Zhang, Fan and
             Liu, Ruiqi and
             Hua, Meng and
             Gao, Zhen and
             Wu, Qingqing and
             Karagiannidis, George K.},
  title   = {Intelligent Metasurface-Enabled Integrated Sensing
             and Communication: Unified Framework and Key Technologies},
  journal = {IEEE Wireless Communications},
  year    = {2026},
  volume  = {33},
  number  = {1},
  pages   = {216--223},
  month   = feb,
  doi     = {10.1109/MWC.2025.3627576}
}

@article{Gholami2026,
  author  = {Gholami, Mehdi and
             Khajavi, Soheil and
             Neshat, Mohammad and
             Tewes, Simon and
             Sezgin, Aydin},
  title   = {Realization of Reconfigurable Intelligent Surfaces
             With Space--Time-Coded Metasurfaces},
  journal = {IEEE Transactions on Antennas and Propagation},
  year    = {2026},
  volume  = {74},
  number  = {5},
  pages   = {4281--4291},
  month   = may,
  doi     = {10.1109/TAP.2026.3660044}
}

@book{StoicaMoses2005,
  author    = {Stoica, Petre and Moses, Randolph L.},
  title     = {Spectral Analysis of Signals},
  publisher = {Prentice Hall},
  address   = {Upper Saddle River, NJ, USA},
  year      = {2005}
}

@article{Ozdogan2020,
  author  = {\"Ozdogan, \"Ozgecan and Bj\"ornson, Emil and Larsson, Erik G.},
  title   = {Intelligent Reflecting Surfaces: Physics, Propagation, and Pathloss Modeling},
  journal = {IEEE Wireless Commun. Lett.},
  volume  = {9},
  number  = {5},
  pages   = {581--585},
  month   = may,
  year    = {2020}
}

@article{Coleri2002,
  author  = {Sinem Coleri and Mustafa Ergen and Anuj Puri and Ahmad Bahai},
  title   = {Channel Estimation Techniques Based on Pilot Arrangement in {OFDM} Systems},
  journal = {IEEE Transactions on Broadcasting},
  year    = {2002},
  volume  = {48},
  number  = {3},
  pages   = {223--229},
  month   = sep,
  doi     = {10.1109/TBC.2002.804034}
}

@article{HuaSarkar1990,
  author  = {Yingbo Hua and Tapan K. Sarkar},
  title   = {Matrix Pencil Method for Estimating Parameters of Exponentially Damped/Undamped Sinusoids in Noise},
  journal = {IEEE Transactions on Acoustics, Speech, and Signal Processing},
  volume  = {38},
  number  = {5},
  pages   = {814--824},
  month   = may,
  year    = {1990},
  doi     = {10.1109/29.56027}
}

@book{anderson1973,
  title={Network Analysis and Synthesis: A Modern Systems Theory Approach},
  author={Anderson, Brian D. O. and Vongpanitlerd, Sumeth},
  year={1973},
  publisher={Prentice-Hall},
  address={Englewood Cliffs, NJ},
  isbn={9780136110538}
}

@book{Felsen1973,
  author    = {Leopold B. Felsen and Nathan Marcuvitz},
  title     = {Radiation and Scattering of Waves},
  publisher = {Prentice-Hall},
  address   = {Englewood Cliffs, NJ},
  year      = {1973}
}

@ARTICLE{Hu2025,
  author={Hu, Changcheng and Zhang, Ruoyu and Wang, Jingqi and Sima, Boyu and Kang, Wei},
  journal={IEEE Sensors Letters}, 
  title={Reconfigurable Coding Design for Transmissive {RIS}-Aided {DOA} Estimation by Single Receiving Sensor}, 
  year={2025},
  volume={9},
  number={7},
  pages={1-4},
  doi={10.1109/LSENS.2025.3573700}}

@article{Chen2026,
  author  = {Chen, Baojun and
             Geng, Zhongzheng and
             Yuan, Zhenglin and
             Zhang, Jie and
             Ju, Yanjie},
  title   = {Direction-of-Arrival Estimation Method for
             Space--Time Coding Metasurfaces Based on
             Harmonic-Domain Complex Correlation},
  journal = {Advanced Theory and Simulations},
  year    = {2026},
  volume  = {9},
  number  = {8},
  pages   = {e70496},
  month   = aug,
  doi     = {10.1002/adts.70496},
}

@ARTICLE{Wang2020,
  author={Wang, Peilan and Fang, Jun and Duan, Huiping and Li, Hongbin},
  journal={IEEE Signal Processing Letters}, 
  title={Compressed Channel Estimation for Intelligent Reflecting Surface-Assisted Millimeter Wave Systems}, 
  year={2020},
  volume={27},
  number={},
  pages={905-909},
  doi={10.1109/LSP.2020.2998357}}

@inproceedings{Rains2023,
  title = {Fully-Addressable Varactor-Based Reflecting Metasurface with Dual-Linear Polarisation for Low Power Reconfigurable Intelligent Surfaces},
  booktitle = {2023 17th {{European Conference}} on {{Antennas}} and {{Propagation}} ({{EuCAP}})},
  author = {Rains, James and Kazim, Jalil Ur Rehman and Tukmanov, Anvar and Zhang, Lei and Abbasi, Qammer and Imran, Muhammad},
  year = 2023,
  month = mar,
  pages = {1--5},
  publisher = {IEEE},
  address = {Florence, Italy},
  doi = {10.23919/EuCAP57121.2023.10133408},
  urldate = {2026-09-03},
  copyright = {https://doi.org/10.15223/policy-029}
}

\end{document}